# Characterisation of microstructural creep, strain rate and temperature sensitivity and computational crystal plasticity in Zircaloy-4

Yang Liu[1*], Weifeng Wan[1], Said El Chamaa[1], Mark R. Wenman[1], Catrin M. Davies[2] and Fionn P.E. Dunne[1]

1 Department of Materials, Imperial College London, London SW7 2AZ, UK

2 Department of Mechanical Engineering, Imperial College London, London SW7 2AZ, UK

## Abstract

Crystal-level strain rate sensitivity and temperature sensitivity are investigated in Zircaloy-4 using combined of bending creep test, digital image correlation, electron backscatter detection and thermo-mechanical tensile tests with crystal plasticity modelling. Crystal rate-sensitive properties are extracted from room temperature microscale creep, and temperature sensitivity from thermal polycrystalline responses. Crystal plasticity results show that large microscale creep strain is observed near notch tip increased up to 50% due to cross-slip activation. Grain-level microscale SRS is highly heterogeneous, and its crystallographic sensitivity is dependent on plastic deformation rate and underlying grain-based dislocation slip activation. Pyramidal <c+a> slip and total dislocation pileups contribute to temperature-sensitive texture effect on yielding and strength hardening. A faithful reconstruction of polycrystal and accurate rate-sensitive single-crystal properties are the key to capture multi-scale SRSs.

## 1. Introduction

Zircaloy-4 is a zirconium alloy widely used as fuel cladding within nuclear fission reactors due to its low neutron absorption, excellent corrosion and mechanical properties. Its nominal chemical composition is Zr-1.5 wt% Sn-0.2 wt% Fe-0.1 wt% Cr (Northwood et al., 1975). Within the reactor, the cladding is subjected to cyclic thermo-mechanical loading at temperatures up to 600 K leading to potential plasticity and creep. During these cyclic reactor operations or spent fuel storage conditions of the high burnup Zircaloy fuel rods, Zircaloy is susceptible to embrittlement due to precipitation of hydrides. Resulting phenomenon, called delayed hydride cracking (DHC), threatens the safety and structural integrity of Zircaloy fuel cladding. The stress fields near notches and cracks, or other stress raisers, are known to attract hydrogen and form local concentrations and the potential hydride formations, which may in turn accelerate DHC and the subsequent crack propagation within the Zircaloy cladding tubes, causing hydride rim formation and hydride reorientation (Chan, 2013).

Macroscopic strain rate sensitivity (SRS) of Zircaloy-4 cladding tubes has been investigated (Lee et al., 2001; Link et al., 1998) using tensile tests at different strain rates, suggesting that the alloy could be considered as strain-rate-insensitive (m < ~0.1) at temperatures up to 700 K. At higher temperatures of 873-1073K, stronger SRSs of about 0.1 - 0.25 were observed in incremental strain rate jumping testing, and oxygen content was found to decrease the SRS in this temperature range (Mehrotra and Tangri, 1980). In the temperature range 573 to 1073 K, Hayes and Kassner measured SRSs with range of 0.16 - 0.2 in both Zircaloy-2 and Zircaloy-4 during creep with applied stress between 0.1 to 115 MPa (Hayes and Kassner, 2017). SRS is known to depend upon deformation history such that the instantaneous SRS is considered as a useful measure since it corresponds to a given microstructure and a given material deformation state (Klepaczko and Chiem, 1986). For hexagonal closest packed (HCP) crystals, e.g. titanium alloys, there is considerable evidence that the intrinsic slip-system-based strain rate sensitivity is anisotropic (Zhang et al., 2016a), i.e., SRSs are different in prism, basal and pyramidal slip systems. Recent study on multi-scale SRS demonstrated the importance of 'microscopic characteristic stress' accounting for the transition from micro- to macro-scale rate sensitivity in polycrystal using visco-plastic self-consistent (VPSC) framework (Zecevic et al., 2016). Skippon et al. measured slip system SRSs in Zircaloy-2 using in-situ XRD *elastic* strain measurements to extract basal, prismatic and pyramidal SRSs (Skippon et al., 2019). They were all found to be less than 0.1 at room temperature, similar to other HCP lattice materials, such as titanium alloys (Jun et al., 2016b, 2016a; Zhang and Dunne, 2017). Skippon et al employed a lattice rate sensitivity definition somewhat unusually based on elastic strains, while assuming uniaxial, uniform stress states within individual grains. The lattice SRSs in pyramidal <c+a> and basal <a> systems were found to be higher than that in prismatic <a>, but were all found to be higher than the macroscopically-measured SRS (Skippon et al., 2019), which is also observed in titanium alloys using XRD technique (Xiong et al., 2020). Zircaloy-4 tube manufacturing often leads to macroscopic material anisotropy and strengthening because of the underpinning crystallographic textures formed (Gurao et al., 2014). Macroscopic texture-dependent SRS was reported early in the 1970s for Zircaloy-4 at 555 K (Wiesinger et al., 1968), where SRS along the TD direction could be up to one order of magnitude higher than for the RD or LD directions with duplex texture of $(0001)[10\bar{1}0]$ and $(0001)[11\bar{2}0]$, reflecting the slip-system-based SRS differences discussed (Zhang et al., 2016a).

A number of modelling approaches have been utilized in order to capture strain rate and temperature sensitivity in the Zr alloys. Lee et al. (2001) assumed a power-law flow rule to describe the effect of dynamic strain aging on the macroscopic SRSs between 523-673 K (Lee et al., 2001). The temperature and strain rate effect on the texture evolution and macroscopic mechanical responses were well captured with a VPSC model with dislocation density and twinning hardening rule included (Knezevic et al., 2015). A numerical procedure to capture SRS over a wide range of strain rates for polycrystalline materials has been developed by Knezevic et al. (2016). A multiscale approach to address SRS was investigated using phenomenological VPSC modelling for zirconium impact tests by Vasilev et al. (2020) and Zecevic et al. (2016). The usage of self-consistent model for crystal plasticity generally weaken or ignore the effect of grain interactions in local plastic deformations, especially when geometrical necessary dislocation is needed for accommodating lattice curvature near grain boundaries (Arsenlis and Parks, 1999). Skippon et al. (2019) utilised a similar EVPSC model to compare with the lattice rate sensitivity of Zircaloy-2 using in-situ XRD techniques. Their predicted lattice rate sensitivity was found to be lower than that from XRD measurement, and applied strain rate has negligible effect on the predicted lattice rate sensitivity, which is contrary to the experimental results for the RD direction (Skippon et al., 2019). 3D synchrotron X-ray diffraction data were compared with fully-resolved crystal plasticity (CP) modelling of grain-resolved stresses and rate sensitivity in polycrystalline Zr in Abdolvand et al. (2018), which has advantages of full 3D information of polycrystals.

This paper addresses SRS in Zircaloy-4 but in a quite different way from previous studies. Here notched blocky-$\alpha$ Zircaloy-4 (Tong and Britton, 2017) sample is utilised and digital image correlation (DIC) measurement at grain level and electron backscatter diffraction (EBSD) are applied to obtain spatial development of room-temperature creep strain and stress within a region of interest (ROI) inside microstructure near the notch. In-situ intragranular and intergranular creep strains can be measured in this way because of the blocky (large) grains. The experimentally characterised microstructure is faithfully reproduced in the CP modelling such that microstructure-level creep and SRS distributions could be analysed. The structure of this paper is described as follows. Section 2 describes the experimental methods adopted. Section 3 presents the method of CP, definition of microscale SRS and representative models for experimental tests. Section 4 shows the calibration process of rate and temperature sensitive crystal-level properties compared with in-situ DIC measurement and texture-dependent thermal tests over temperature range of 293K to 623K. Section 5 utilises the extracted properties to analyse microscale creep, SRS and stress and temperature sensitive texture effect. In passing, the capability of the model to capture polycrystalline texture effects is demonstrated. Discussions on multiscale SRS in separate tests and uncertainties of current method are raised in Section 6, followed by conclusions in Section 7.

## 2 Experiments

For obtaining the strain rate sensitive and temperature sensitive material properties for single crystal Zircaloy-4, two set of experiments are performed including three-point creep bending and thermo-mechanical tensile tests, respectively.

### 2.1 Three-point creep bending with DIC and EBSD characterisation

Three-point creep bend testing of a notched sample machined from blocky alpha (large grained) Zircaloy-4 has been carried out as shown schematically in Figure 1(a). Force-controlled loading is applied which contains a substantive force hold period shown in Figure 1(b). Utilising the processing method introduced in Tong and Britton (2017), as-received Zircaloy-4 sample with fine grain structure was heated to 800 °C for 2 weeks to achieve large grains known as 'blocky $\alpha$', and the average grain size increased from ~ 20 µm to ~ 400 µm within the sample. Strong texture is observed in the sample (shown later) where the c-axes of most of blocky $\alpha$ grains are orientated out-of-plane as indicated in Figure 1(a).

The sample surface was firstly machine ground and polished carefully followed by electropolishing under 25 V at -30 °C for 1 minute, which aims to prepare for the EBSD scan. Local strain measurement at the ROI is carried out by optical microscopy based DIC. 1 µm silica particles are homogeneously speckled on surfaces of the sample as shown in Figure 2(a). In-situ optical DIC testing is carried out using a Questar Microscope Lens (QM-100), a QI-click versatile scientific CCD camera, a three-point bend testing rig and a three-axis sliding supporting stage. The three-point bend testing setup applied on a Shimadzu universal testing machine was introduced in previous work (Chen et al., 2017). Questar microscope lens is connected to the camera by a C-mount adaptor with a fibre-optic illuminator supplying coaxial illumination. The lens is mounted on a supporting stage, which allows sliding along X, Y and Z directions, and the camera is supported by a screwed metal cylinder attached to the supporting stage to avoid motion. The Questar microscope with the highest resolution of 1 µm can recognize the sprayed speckles with a size of 1 µm on the sample surface, shown in Figure 2 (b), when the working distance is around 150 mm. The speckle images were captured in situ by the computer-controlled camera during the three-point bend testing. After testing the captured images were used to carry out strain calculation using an in-house MATLAB code.

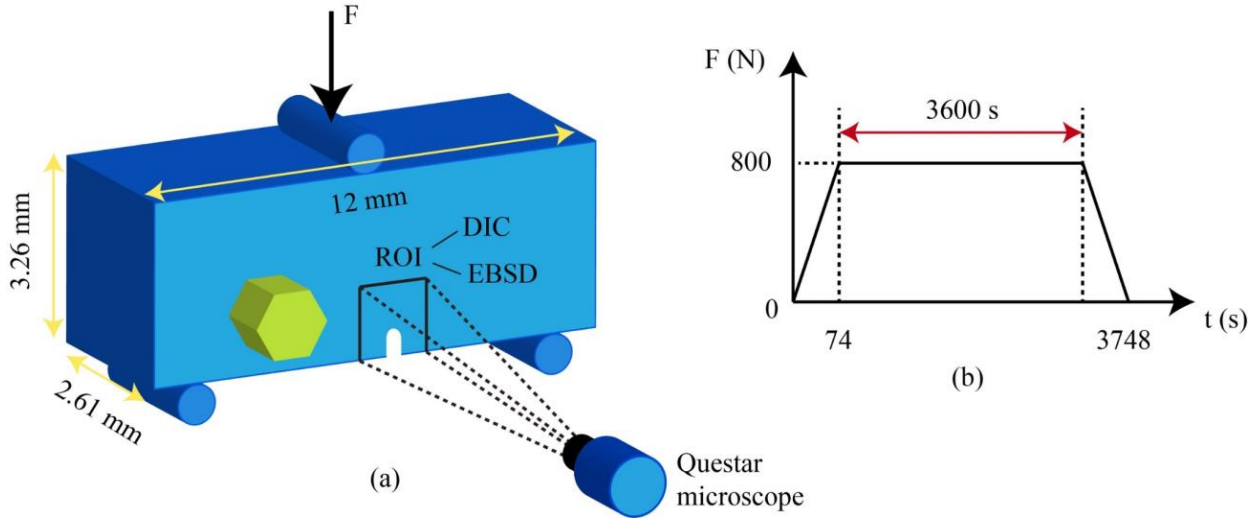

*Figure 1 shows (a) the three-point bending test with the ROI scanned by Questar microscope to extract the strain distribution at grain scale using DIC combined with EBSD (c-axis out-of-plane) and (b) the loading condition at the top of the loading beam. The crystal in (a) indicates schematically the predominant texture of the polycrystal blocky α sample with most grain c-axes out of plane.*

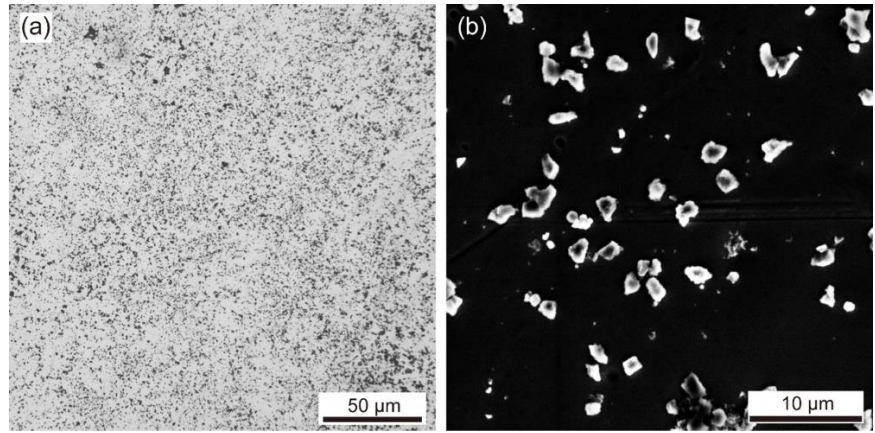

*Figure 2 OM image (a) and SEM image (b) showing the distributed 1 μm silica speckles on the surface of sample A for DIC.*

## 2.2 Thermo-mechanical load

Thermo-mechanical tensile tests were performed on dog bone specimens which were machined as shown in Figure 3 (a), extracted from bulk Zircaloy-4 plate along both rolling direction (RD) and transverse direction (TD) shown in Figure 3 (b). such that crystallographic texture effects can be examined. The normal direction (ND) behaviour is not examined due to the limited thickness of bulk plate. The EBSD scan with inverse pole figure (IPF) along the Z direction (ND) is shown in Figure 7 (c), where the c axes of many grains align approximately parallel to the Z direction. The thermo-mechanical tests were conducted at the same strain rate of 0.01/s at 293 K and 623 K for both RD and TD specimens.

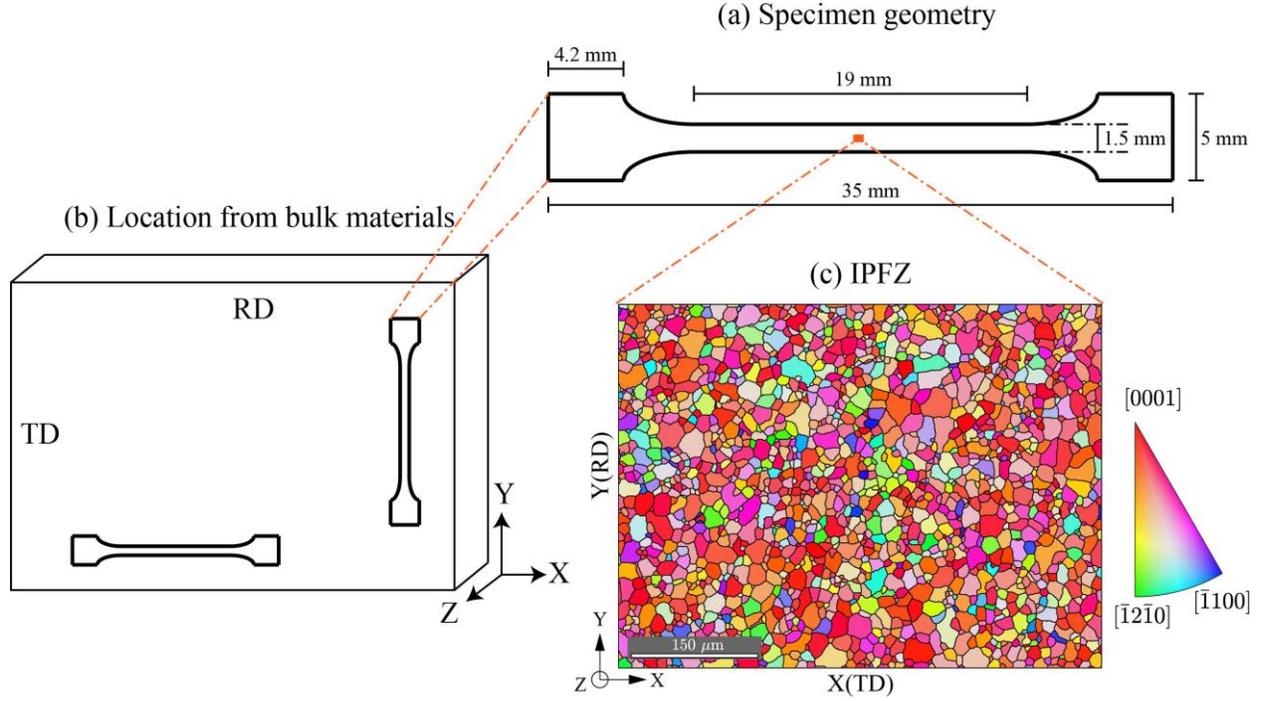

*Figure 3 (a) dog bone specimen geometry extracted from (b) bulk zicaloy-4 sample along RD and TD directions. (c) EBSD scan with and IPFZ colour map and grains reconstructed with 5° misorientations angle limit.*

## 3 Methods

CP theory is introduced in brief and multi-variable function of SRS is defined according to thermal activation mechanism with reconstructed models for ROIs in bending and tensile tests.

### 3.1 CP theory

The CP model used in this study has been demonstrated for HCP polycrystals including the titanium alloys (Dunne et al., 2007; Zhang et al., 2016b), to capture strain rate sensitivity, stress redistribution and creep for a range of thermo-mechanical loading. Here, it is extended for zirconium alloys using temperature and rate dependent CP theory. The total deformation gradient is multiplicatively decomposed into anisotropic elastic, plastic and thermal parts as

$$\mathbf{F} = \mathbf{F}^e \mathbf{F}^p \mathbf{F}^\theta \tag{1}$$

The anisotropic thermal expansion $\mathbf{F}^\theta$ and its rate form are given by,

$$\mathbf{F}^\theta = \mathbf{I} + (\theta - \theta_0)\boldsymbol{\alpha}; \quad \dot{\mathbf{F}}^\theta = \dot{\theta}\boldsymbol{\alpha} \tag{2}$$

where, $\theta$ is the temperature in Kelvin and $\boldsymbol{\alpha}$ is a diagonal tensor in the crystallographic configuration, with diagonal components $\alpha_1 = \alpha_2 = 1.01 \times 10^{-5}$ /K, $\alpha_3 = 5.30 \times 10^{-6}$ /K (Abdolvand, 2019). $\boldsymbol{\alpha}$ tensor is rotated appropriately to the global configuration using EBSD measured orientation data.

The plastic velocity gradient $\mathbf{L}^p$ is related to the shear slip rate

$$\mathbf{L}^p = \dot{\mathbf{F}}^p(\mathbf{F}^p)^{-1} = \sum_s \dot{\gamma}^s \mathbf{m}^s \otimes \mathbf{n}^s \qquad (3)$$

where, $\dot{\gamma}^s$ is the shear strain rate of the $s^{th}$ slip system with slip normal $\mathbf{n}^s$ and slip direction $\mathbf{m}^s$. Corresponding Schmid factor $M^s$ of the $s^{th}$ slip system is defined as $(\mathbf{l}\cdot\mathbf{n}^s)(\mathbf{l}\cdot\mathbf{m}^s)$ and $\mathbf{l}$ is the remote load direction.

The dislocation slip along the $s^{th}$ slip system is based on the thermally activated escape of pinned dislocations to overcome the energy barrier resulting from obstacles. Slip system shear strain rate $\dot{\gamma}^s$ is controlled by the averaged dislocation velocity given by Cottrell and Dexter (Cottrell and Dexter, 1954)

$$\dot{\gamma}^s = \rho_m b^s v_g \qquad (4)$$

where $b^s$ is the Burgers vector magnitude of this slip system, $\rho_m$ the mobile dislocation density and $v_g$ the average dislocation glide velocity. Considering both forward and backward dislocation jumps (Dunne et al., 2007), the shear strain rate that results when the resolved shear stress $\tau^s$ exceeds its critical value $\tau_c^s$ is

$$\dot{\gamma}^s = \rho_m \omega (b^s)^2 \exp\left(-\frac{\Delta F}{k\theta}\right) \sinh\left(\frac{\Delta V}{k\theta}(\tau^s - \tau_c^s)\right) \qquad (5)$$

where $\omega$ is the dislocation jump frequency. The thermal activation energy $\Delta F$ and activation volume $\Delta V$ are the two crucial parameters controlling the rate sensitivity of dislocation slip, and $k$ is the Boltzmann constant. The 'sinh' function ensures the description of strain rate sensitivity is appropriate at both low and high strain rate, discussed elsewhere (Zhang and Dunne, 2017). The relationships between slip system critical resolved shear stresses in Zircaloy-4 have been investigated and quantified independently using micro-cantilever testing to give $\tau_c^{basal} = 1.33\tau_c^{prism}$; $\tau_c^{pyramidal} = 3.48\tau_c^{prism}$ (Gong et al., 2015).

Based on the Taylor hardening model (Taylor, 1934), the critical resolved shear stress (CRSS) evolution is related to the total dislocation density, comprising statistically stored (SSD) and geometrically-necessary dislocations (GND)

$$\tau_c^s = \tau_{c,0}^s + Gb^s\sqrt{\rho_{SSD} + \rho_{GND}} \qquad (6)$$

where $\tau_{c,0}^s$ is the initial slip system CRSS and $G$ is shear modulus. The GND densities are considered to accommodate the crystal lattice curvatures and lead to localised intragranular strain hardening. Following (Dunne et al., 2007; Nye, 1953), the density of GNDs is computed by relating the curl of the plastic deformation gradient to the Nye tensor.

The evolution rate of $\rho_{SSD}$ is based on the plastic strain rate $\dot{\bar{\varepsilon}}^p$,

$$\rho_{SSD} = \gamma_s \dot{\bar{\varepsilon}}^p \tag{7}$$

where $\dot{\bar{\varepsilon}}^p = \sqrt{2/3(\dot{\varepsilon}_{ij}^p \dot{\varepsilon}_{ij}^p)}$ and $\gamma_s$ is the hardening coefficient for statistically stored dislocation density, which controls the strength increase during the strain hardening process.

### 3.2 SRS function

The strain rate sensitivity (SRS) $m$ is generally considered as a function of strain, strain rate, crystallography and temperature,

$$m = f(\varepsilon, \dot{\varepsilon}, M^s, \theta) \tag{8}$$

The conventional definition of macroscale SRS $m$ is

$$m = \frac{\partial \ln \sigma}{\partial \ln \dot{\varepsilon}^p} \tag{9}$$

where $\sigma$ is the magnitude of uniaxial stress and $\dot{\varepsilon}^p$ is the plastic strain rate. This is generally measured macroscopically by multiple monotonic tensile tests under different strain rates and is anticipated to be dependent on both the applied strain influencing underpinning evolution of microstructural quantities including dislocation density, lattice curvature and strain rate reflecting thermally activated events.

The SRS for a single crystal undergoing slip was investigated in (Zhang and Dunne, 2017). They showed that the crystal SRS is not an intrinsic property but rather, depends upon crystal orientation with respect to loading direction and other intrinsic slip system properties including slip strength $\tau_c^s$, activation energy $\Delta F$, and corresponding activation volume $\Delta V$. It follows the dislocation glide velocity (Evans and Rawlings, 1969; Hirth and Nix, 1969), crystal slip rate $\dot{\gamma}^s$ and their relationship via Schmid factor, $M^s$ to the equivalent macroscopic direct strain rate $\dot{\varepsilon}^p$, given by

$$\dot{\varepsilon}^p = \rho_m b^s v_g M^s = \dot{\gamma}^s M^s \tag{10}$$

The corresponding single crystal uniaxial loading direction stress $\sigma$ was derived in (Zhang and Dunne, 2017) from Eq. (5) to be

$$\sigma = \frac{\tau^s}{M^s} = \frac{1}{M^s}\left[\frac{kT}{\Delta V}\sinh^{-1}\left(\frac{\dot{\varepsilon}^p}{\eta \exp\left(-\frac{\Delta F}{k\theta}\right)M^s}\right) + \tau_c^s\right] \tag{11}$$

where $\eta = \rho_m \omega (b^s)^2$. Hence the SRS of the single crystal was obtained from Eq. (7) to be

$$m = \frac{\dot{\varepsilon}^p}{\sqrt{\left(\eta \exp\left(-\frac{\Delta F}{kT}\right)M^s\right)^2 + (\dot{\varepsilon}^p)^2}\left[\frac{\dot{\varepsilon}^p}{\eta \exp\left(-\frac{\Delta F}{k\theta}\right)M^s} + \frac{\Delta V}{kT}\tau_c^s\right]} \tag{12}$$

Hence, Eq. (10) shows an analytical explicit formula of SRS as a function of Schmid factor $M^s$, plastic strain rate $\dot{\varepsilon}^p$ and temperature $\theta$. The effect of strain $\varepsilon$ is implicitly incorporated in $\tau_c^s$ through dislocation density-based strain hardening mechanism where the evolution of $\rho_{GND}$ depends on local plastic strain gradient curl($F^p$) and $\rho_{SSD}$ directly depends on plastic strain accumulation ($\bar{\varepsilon}^p$). The crystal SRS $m$ is explicitly shown as not a basic crystal property but in fact a function depends upon temperature, crystal orientation (via Schmid factor $M^s$) and key slip system properties including $\tau_c^s$, thermally activated $\Delta F$ and $\Delta V$ which govern the slip rate in Eq. (5), in addition to strain rate and strain when slip system hardening occurs.

### 3.3 Representative Model

Finite element (FE) models are reconstructed based on the microstructural information from EBSD scans of Zircaloy-4 samples. Blocky-$\alpha$ grains and as-received equiaxed grains are obtained and representative volume elements (RVEs) are reconstructed in the ROIs for bending and tensile tests, respectively.

### 3.3.1 Bending test model

The sample free-surface EBSD characterization has been utilized to develop a faithful microstructural model reproduction in the ROI shown in Figure 1(a), capturing both free-surface grain morphology and crystallographic orientations. Figure 4 shows the FE geometrical microstructure model together with the grain crystallographic data obtained from EBSD with a scan resolution of 2 µm. The grain boundaries were reconstructed based on 5° misorientation angle limit, resulting in 329 grains, on the ROI, with orientation averaged within individual grain. Grain boundary geometry is imported into the commercial FE platform ABAQUS using in-house python scripts and smoothed at each grain boundary to eliminate the staircase effect. The EBSD scan shows a textured sample with predominant c-axis orientation out-of-plane, reflected by the ODF in Figu(b). Primary bending stress loading is therefore perpendicular to the c-axes of most grains, so that prismatic slip activation is anticipated to be predominant above other slip systems. The FE mesh with 135175 quadratic hexahedral elements (C3D20) is shown for explicit grain structures in Figure 4 *Figu*(c).

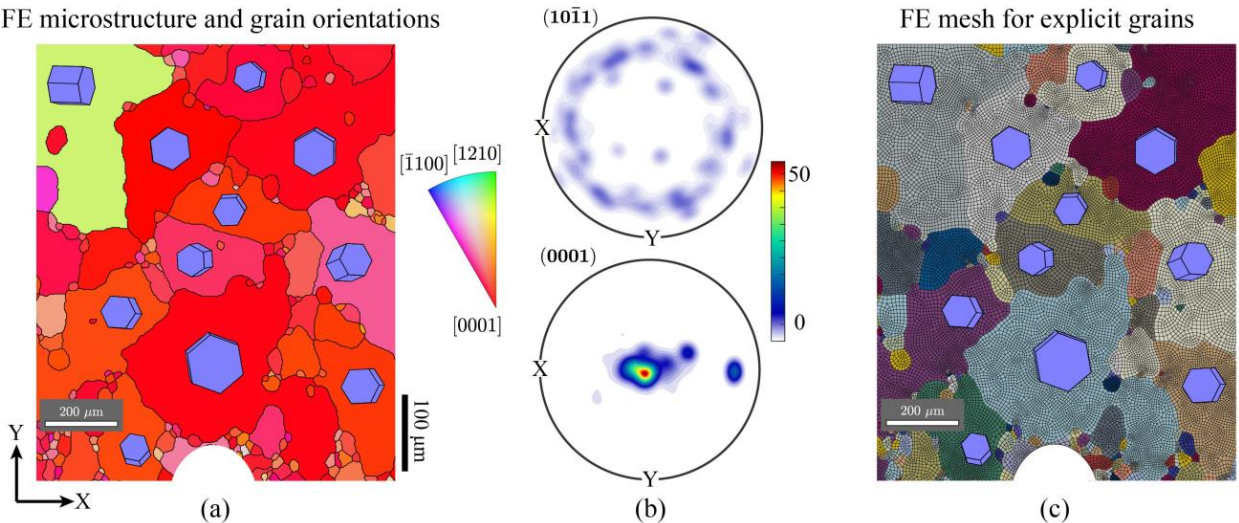

*Figure 4 shows the microstructure reconstruction of the ROI in sample A: (a) Grain geometry and orientations inputs in ABAQUS, from EBSD scan and MTEX reconstruction (IPFZ: out-of-plane c axis in unrotated crystal configuration.); (b) Texture shown in pole figures on crystal planes* $(0001)$ *and* $(10\bar{1}1)$*; (c) finite element mesh for the explicitly represented grain structure.*

Figure 5 (a) shows a schematic of the notched beam subject to three-point bending conditions, with constraint boundary and loading conditions shown. The 800 N force is applied on the top surface over a contact area of 2.61×0.8 mm$^2$. ROI for both EBSD and optical DIC measurement is chosen near the notch. To increase computational efficiency, a sub-model is used shown in Figure 5 (b) which is marked by the white line on the surface of whole sample in Figure 5 (a), replacing full beam utilised for CP modelling. The details of boundary conditions selected are discussed in Appendix A. The surface geometry is extruded 50 μm in the z-direction maintaining plane stress conditions on the free surface. The sub-model contains the ROI, which is the polycrystal region explicitly reproduced from the experiment in Figure 4. Crystal elasto-plasticity behaviour is assigned to the explicitly represented grains in ROI (the EBSD scanned area), and outside of this region within the sub-model, is assigned utilising Mises plasticity behaviour for Zircaloy-4 from previous work (Wilson et al., 2019). In order to demonstrate the validity of non-uniform load applied to the sub-model, a full beam bending is modelled assuming Mises's plasticity behaviour for the full and sub-models and the stress component $\sigma_{xx}$ is shown in Figure 5 (c), which shows nearly same stress state compared with non-uniform stress load on the sub-model. Furthermore, stress $\sigma_{xx}$ along path A-A' is compared between full beam model and sub-model in Figure 5 (d) at the end of load up in load history of Figure 1 (b), which shows good agreement and demonstrates that the loading condition applied in sub-model is equivalent to the three-point bending load applied in full model.

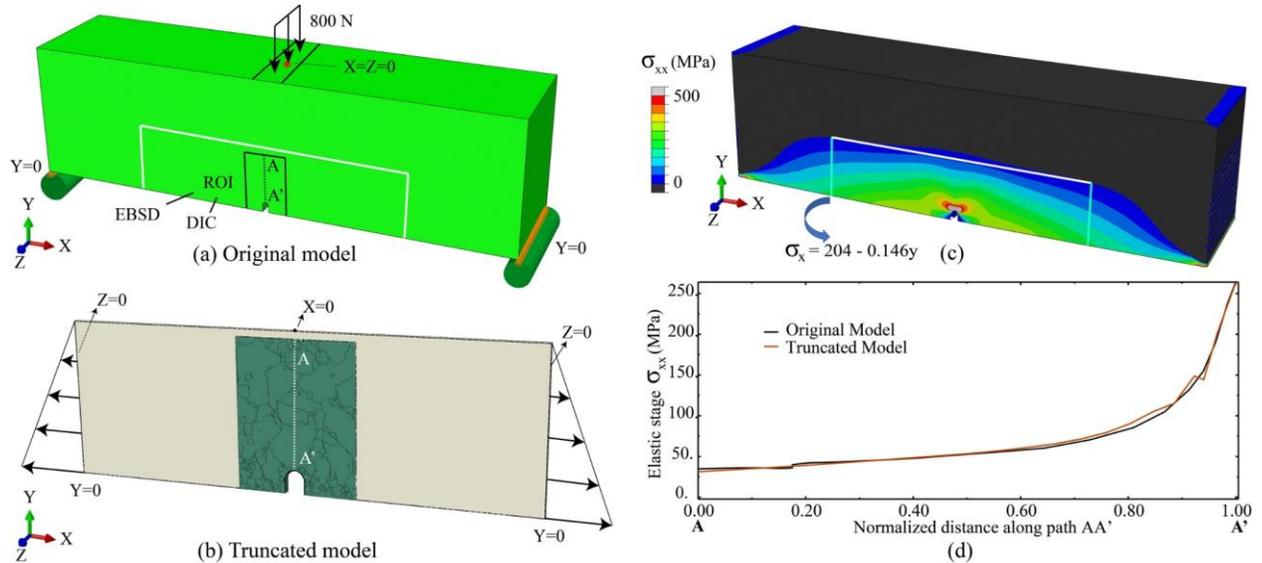

*Figure 5 shows (a) the simplified three-point bending test and the ROI of EBSD and DIC near the notch and (b) a truncated region of model reduction (path AA' is chosen for stress comparison). (c) Stress component $\sigma_{xx}$ distribution on the surface of the notch shows the same linear stress reduction along Y-axis*

*at the edge of truncated model. (d) Stress component σ$_{xx}$ is compared between original and truncated model along path A-A' at end of load up in Figure 1 (b).*

### 3.3.2 Tensile RVE model

The grain-based RVE is reconstructed for modelling with grain boundaries being determined based on experimental EBSD scan of as-received bulk polycrystalline Zircaloy-4 sample, shown in Figure 3 (c). Regarding mesh type and mesh density in the RVE, the quadratic hexahedral element type C3D20 with mesh density shown in Figure 6 (a) provides sufficient precision to capture the polycrystalline and local single-crystal responses at satisfactory speed, which is shown in detail in Appendix B. It is noted that conformal grain mesh generation with hexahedral elements could lead to relatively better local grain responses (Barrett et al., 2018; Feather et al., 2020; Knezevic et al., 2014). Figure 6 (b) shows the grain statistics from experimental measurement, including aspect ratio, size distribution, sphericity, crystallographic orientation and misorientation distribution (within angle bin of 10°). From the pole figures in Figure 6 (b), the c-axes of most grains align in the Z-X plane and the loading directions along RT and TD are anticipated to have a strong effect on the resulting behaviours, leading to a texture-dependent elasto-plastic response. The geometric information is imported via NEPER (Quey et al., 2011) to generate a model polycrystal and the crystallographic data are assigned by enforcing the misorientation distribution using the misorientation probability assignment method (MPAM) (Deka et al., 2006). The reconstructed 3D model microstructure includes 300 grains, and the non-conformal grain boundary has been investigated before to show limited effect on local mechanical behaviour (Chen, 2018).

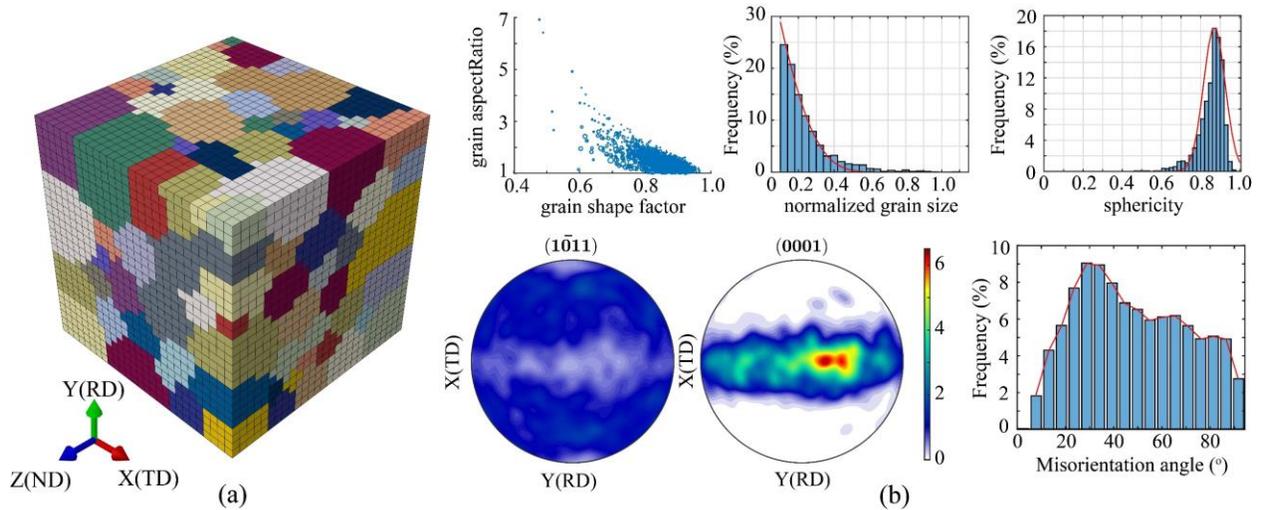

*Figure 6 (a) reconstructed microstructure including 300 grains; (b) statistical information of grain geometry and grain orientation distribution in pole figure.*

### 4. Crystal properties extraction

Based on the CP theory and reconstructed models for the ROIs in creep bending and tensile tests, the rate and temperature-sensitive single crystal properties are extracted for Zircaloy-4 using CP modelling in comparison with experimental data from in-situ DIC measurement and polycrystalline stress-strain responses, respectively.

## 4.1 Rate-sensitive property

CP modelling is used to determine the rate-sensitive properties compared with grain-level in-situ DIC-measured creep strain evolution. The CRSS values of prismatic and basal are set the same as properties has been calibrated (Wilson and Dunne, 2019) prior to any hardening due to dislocation density evolution. Several crystal properties in thermal activation theory of Eq.(5) were determined including Burgers vector $b^s$, dislocation jump frequency $\omega$ and initial mobile dislocation $\rho_m$ (Gong et al., 2015; Wilson and Dunne, 2019), except the two most important ones, thermal activation energy $\Delta F$ and thermal activation volume $\Delta V$, which controls the single-crystal and microstructural rate sensitivity (Zhang and Dunne, 2017).

In order to obtain $\Delta F$ and $\Delta V$, two parallel spots labelled 1 and 2 in the microstructure shown in Figure 7 (a) are chosen with different crystallographic orientations to extract the crystal properties which give best match between CP and experimental data (least square optimisation). To eliminate the errors from extreme values in local pixels of DIC measurement, a non-local definition of experimental creep strain for spot 1 and 2 is used, which considers the average of 4 pixels (radius of 6 µm) in the DIC results. Similarly, the CP result of the same spot is averaged within a circle of 6 µm radius.

The activation energy $\Delta F$ was determined in the range of 0.22-2.89 eV (Adamson, 1977; Holmes, 1964; Matsuo, 1987) and activation volume $\Delta V$ of 10-300 $b^3$ (Cuniberti and Picasso, 2001; Kombaiah and Murty, 2015; Sun Ig Hong et al., 1984a) at temperature range of 293K to 623K in this study. During rate-sensitive property extraction, spread sets of ($\Delta F$, $\Delta V$) within these ranges are considered to firstly capture the creep strain $\varepsilon_{xx}$ evolution of spot 1 along remote stress direction (X-axis). Then the matching set is secondly checked compared with $\varepsilon_{xx}$ evolution of spot 2. This process is done for all spread set and only one set of ($\Delta F$, $\Delta V$) = (0.32 eV, 20.93 $b^3$) is extracted to best match both creep strain evolutions of these two spots shown in Figure (b). It is noted that the irreversible plastic strains at the unload state are captured well, which means that the dislocation-driven thermal activation mechanism and corresponding properties ($\Delta F$, $\Delta V$) are sufficient to describe the microstructural creep evolution at different crystallographic orientations.

To better validate the rate-sensitive properties extracted, the line profile result from CP is compared with DIC measurement along path B-B' at different stress hold steps and the unload state. The observed behaviour near the grain boundaries (indicated by the vertical broken lines with higher misorientations of 11.5° and 8°, next to spot 1 and 2, respectively) is captured well by the extracted properties.

## 4.2 Temperature-sensitive slip strength

Zircaloy-4 application in nuclear reactors involves service conditions of thermo-mechanical loading at high temperatures reaching ~650 K. This section addresses the temperature sensitivity of single-crystal properties over the range of 293K to 623K and its validation against experimental data, mainly the temperature sensitivity of slip strengths. The anisotropic elastic stiffness constants were extracted approximately linear functions of temperature over range of 293K to 623K (Fisher and Renken, 1964). It is noted that the extracted activation energy $\Delta F$ and corresponding activation volume $\Delta V$ from thermally activated dislocation slip were demonstrated experimentally with small variation over the temperature range considered (Cuniberti and Picasso, 2001; Kombaiah and Murty, 2015). Thus, they remain the same while obtaining temperature sensitive properties.

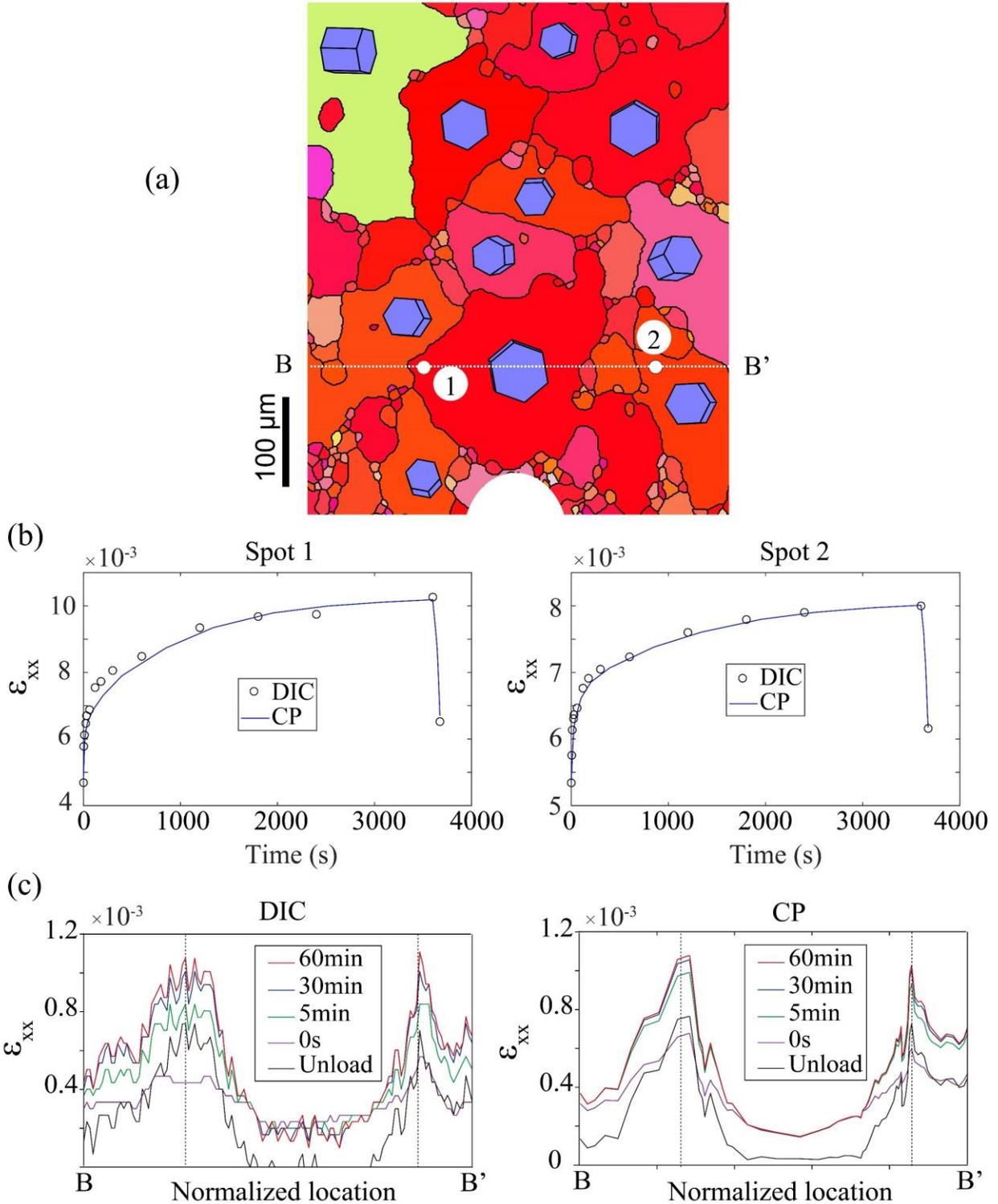

*Figure 7 (a) calibration spots chosen from ROI and a path B-B' crossing the specimen; (b) spot-wise calibration for whole loading history; (c) strain comparison along path B-B' between DIC-measurement*

*and CP modelling at different stress holding period and unload state. The vertical broken lines are the locations of grain boundaries with relatively higher misorientations, next to spot 1 and 2, respectively.*

Micro-pillar tests on zircaloy-4 have enabled the extraction of temperature-dependent slip strength (CRSS) of the basal slip system and an exponential function was found to best represent the temperature dependence (Wang et al., 2019). Thus, the exponential function is utilized for obtaining the temperature dependence of prism slip strength, $\tau_c^{prism} = f(\eta, \zeta) = \eta \exp(\zeta/\theta)$, prior to any hardening behaviour, and $\tau_c^{prism}$ at room temperature is consistent with that in Section 4.1. The ratios between the differing prism, basal, and pyramidal slip system strengths determined in Gong et al. (2015) are utilized and assumed to be independent of temperature over the range considered.

The reconstructed microstructural RVE in Figure 6 (a) of a mid-section of the dog bone sample in Figure 3 is under uniaxial tensile conditions, where $U_x = 0$, $U_y = 0$ and $U_z = 0$ are applied on surfaces $X = 0$, $Y = 0$ and $Z = 0$ respectively, with corresponding opposite faces free, and strain-loaded along *TD (X)* direction. The distribution of stress component $\sigma_{xx}$ is shown in Figure 8 (a) at 623 K with average stress level of 250 MPa. A similar two-step iteration used in Section 4.1 is employed here for obtaining $(\eta, \zeta) = (26.3, 514.7)$ while comparing the CP results to experimental tensile data. Good agreement is achieved at both 293K and 623 K with strain magnitude of 0.02 and strain rate of 0.01/s along *TD* direction shown in Figure 8 (b).

To demonstrate the validity of the extracted temperature-sensitive material properties for other temperatures within range of 293K to 623K and other quasi-static strain rates, the CP polycrystal model in Figure 6 (a) has been subjected to straining at various temperatures to obtain the yield stresses which are compared with the experimental values at strain rate of 0.001/s (Kumar et al., 2018) and the results are shown in Figure 8 (c). It is noted that the microstructure reported in (Kumar et al., 2018) exhibits similar grain size and texture as the current Zircaloy-4 sample and its basal pole is also preferably oriented for the ND direction. Here, the yield stress is chosen as that at the offset 0.2% strain. To further check its validity at different strain rates, the CP modelling is also conducted at another strain rate of $3.3 \times 10^{-5}$/s to compare with a different set of tensile tests conducted for Zircaloy-4 with similar chemical composition (Derep et al., 1980), and shows reasonable agreement in Figure 8 (c).

In conclusion, the rate and temperature-sensitive crystal properties are extracted and validated using CP modelling compared with experimental results over temperature range of 293K to 623K and strain rate range of $3.3 \times 10^{-5}$/s to 0.01/s. The obtained properties are shown in Table 1.

## 5. Results

Upon extracted single-crystal properties, the micromechanical and polycrystalline responses are analysed, including creep strain, corresponding microscale SRS, and microstructural effect, revealing the role of micro-slip rate/accumulation, local crystallography and dislocation accumulation on crystal-level heterogeneous SRS and temperature-dependent texture effect.

### 5.1 Large microscale creep strain

Zircaloy-4 has been considered to be strain-rate insensitive at temperature up to 700 K (Lee et al., 2001; Link et al., 1998) and a corresponding high activation volume is argued on the basis of a bulk Zircaloy-4 (Kombaiah and Murty, 2015; Sun Ig Hong et al., 1984b). This generally indicates that limited rate-sensitive creep will occur in Zircaloy-4. However, the room-temperature DIC measurements and CP results in this study exhibit very significant time-dependent strain evolution, which demonstrate strong creep response at microscale. Quantitatively shown in Figure 9 (a), the beginning strain $\varepsilon_{xx}$ of spot 3 near notch tip in Figure 9 (b) (at the end of load hold) is 1.2% and a total creep strain accumulation of 0.6% is achieved after stress holding of only 1 h, exhibiting 50% strain increase during creep. Practically, in-service Zircaloy-4 experience much higher longer stress holding period at higher temperatures, leading more significant creep. The high accumulated microscale creep strain especially at locations of stress raisers, would potentially leads to significant fatigue crack nucleation and stress redistribution during thermo-mechanical cycles, which needs to be considered in evaluating the structural integrity of Zircaloy cladding inside nuclear reactors.

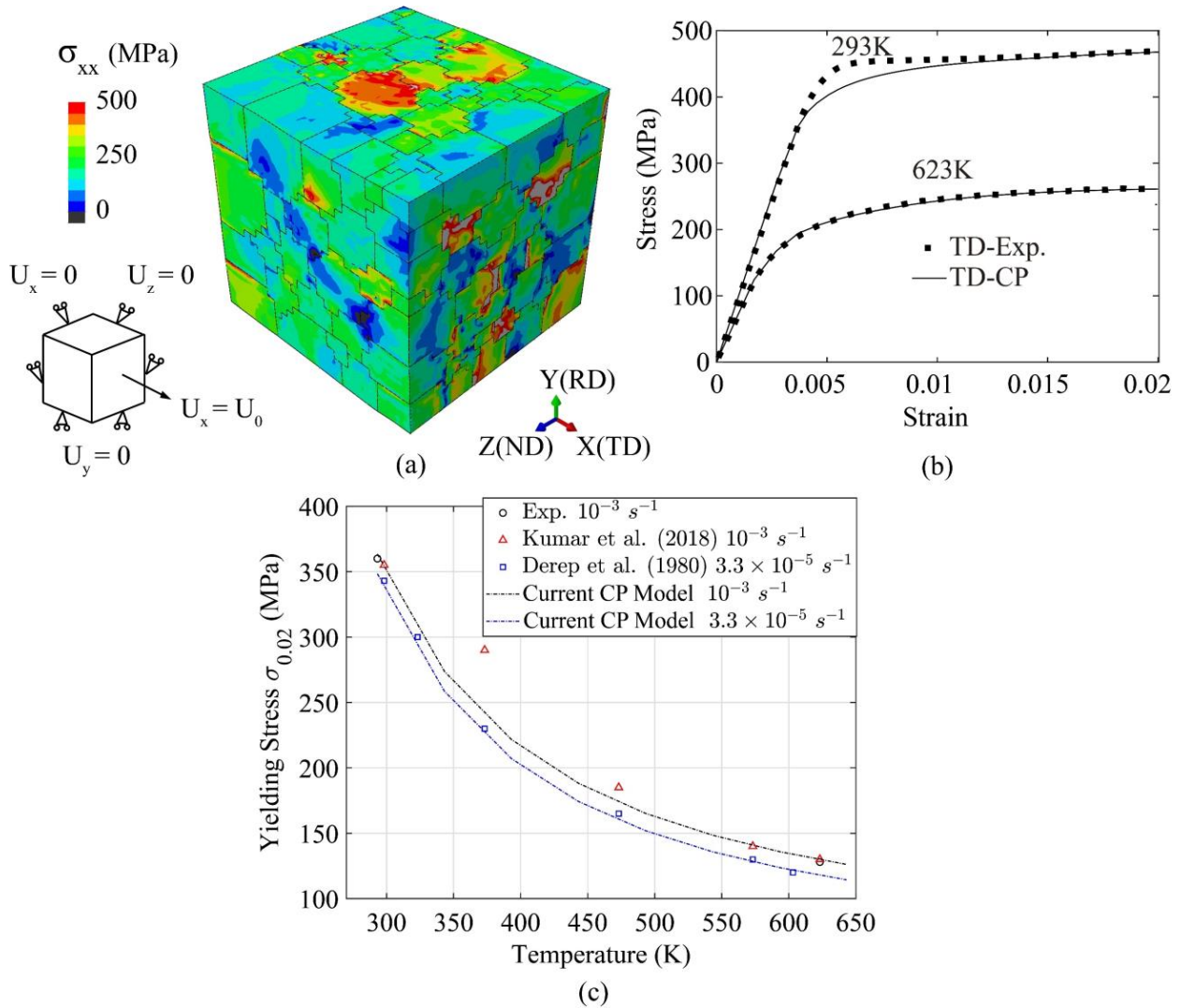

*Figure 8 (a) stress $\sigma_{xx}$ state of RVE at the end of strain 0.02 tensile at 623 K. (b) stress-strain comparison between tensile experiments and CP results. (a) and (b) are at 0.01/s strain rate loaded along X(TD)-axis. (c) extracted temperature-sensitive yielding stresses $\sigma_{0.02}$ in comparison with Zircaloy-4 tensile tests at strain rates of $1\times10^{-3}$/s and $3.3\times10^{-5}$/s.*

*Table 1 Extracted elasto-plastic single-crystal properties for Zircaloy-4*

| Crystal property | | Units | Quantities |
|---|---|---|---|
| Elastic | $E_1$ | MPa | $E_1 = -0.0755\theta + 120.4411$ |
| | $E_3$ | MPa | $E_3 = -0.0327\theta + 132.8621$ |
| | $G_{12}$ | MPa | $G_{12} = -0.0233\theta + 38.8367$ |
| | $v_{12}$ | | $v_{12} = 3.4273\times10^{-4}\theta + 0.3002$ |
| | $v_{13}$ | | $v_{13} = -9.1182\times10^{-5}\theta + 0.2642$ |
| Plastic | $\rho_m$ | $\mu m^{-2}$ | 0.01 |
| | $\omega$ | Hz | $1.0\times10^{11}$ |
| | $b$ | $\mu m$ | $3.2\times10^{-4}$ |
| | $k$ | $J \cdot K^{-1}$ | $1.381\times10^{-23}$ |
| | $\Delta V$ | $b^3$ | 20.93 |
| | $\tau_c^\alpha$ | MPa | $\tau_c^{prism} = 26.3\exp(514.7/\theta)$ |
| | $\Delta F^\alpha$ | J | $5.12697 \times 10^{-20}$ |
| | $\gamma_s$ | $\mu m^{-2}$ | 20 |

Individual slip system activations are shown within blocky-$\alpha$ grain near notch tip in Figure 9 (c). The slip activation is found to be purely prismatic, with the main slip activation on slip system $(0\bar{1}10)[\bar{2}110]$ ($\mathbf{n}_1 \otimes \mathbf{m}_1$) and some modest activation near the notch on system $(10\bar{1}0)[\bar{1}\bar{2}10]$ ($\mathbf{n}_2 \otimes \mathbf{m}_2$). Both slip activations contribute to the creep strain accumulation in spot 3, which demonstrates that multi-slip activation and potential slip interactions are one of the major causes of high creep strain accumulation. There is virtually no activation on the third prism system $(1\bar{1}00)[11\bar{2}0]$ ($\mathbf{n}_3 \otimes \mathbf{m}_3$), which has low Schmid factor with respect to the remote load and hence is not well-oriented for slip.

Figure 10 shows the spatial distribution of strain component $\varepsilon_{xx}$ obtained from the experimental DIC measurements (labelled Exp. DIC) at the time points in the loading history (during hold time) shown, and at the unloaded state. With increasing stress-hold time from 30 s to 20 min, the strain develops into the sample and forms a 'butterfly wings' pattern. Asymmetry in the strain field develops which results from the local grain morphologies, their interaction with neighbouring grains which are adjacent to the dominating grain marked with white line in Figure 9 (b) at the sample notch tip. The strain evolutions show a progressive decrease in rate with holding time, reflecting a creep deformation process and suggesting dislocation hardening. Hence the strain increases during the hold between 20 to 60 min with progressively diminishing creep rate as it approaches the steady state secondary creep rate.

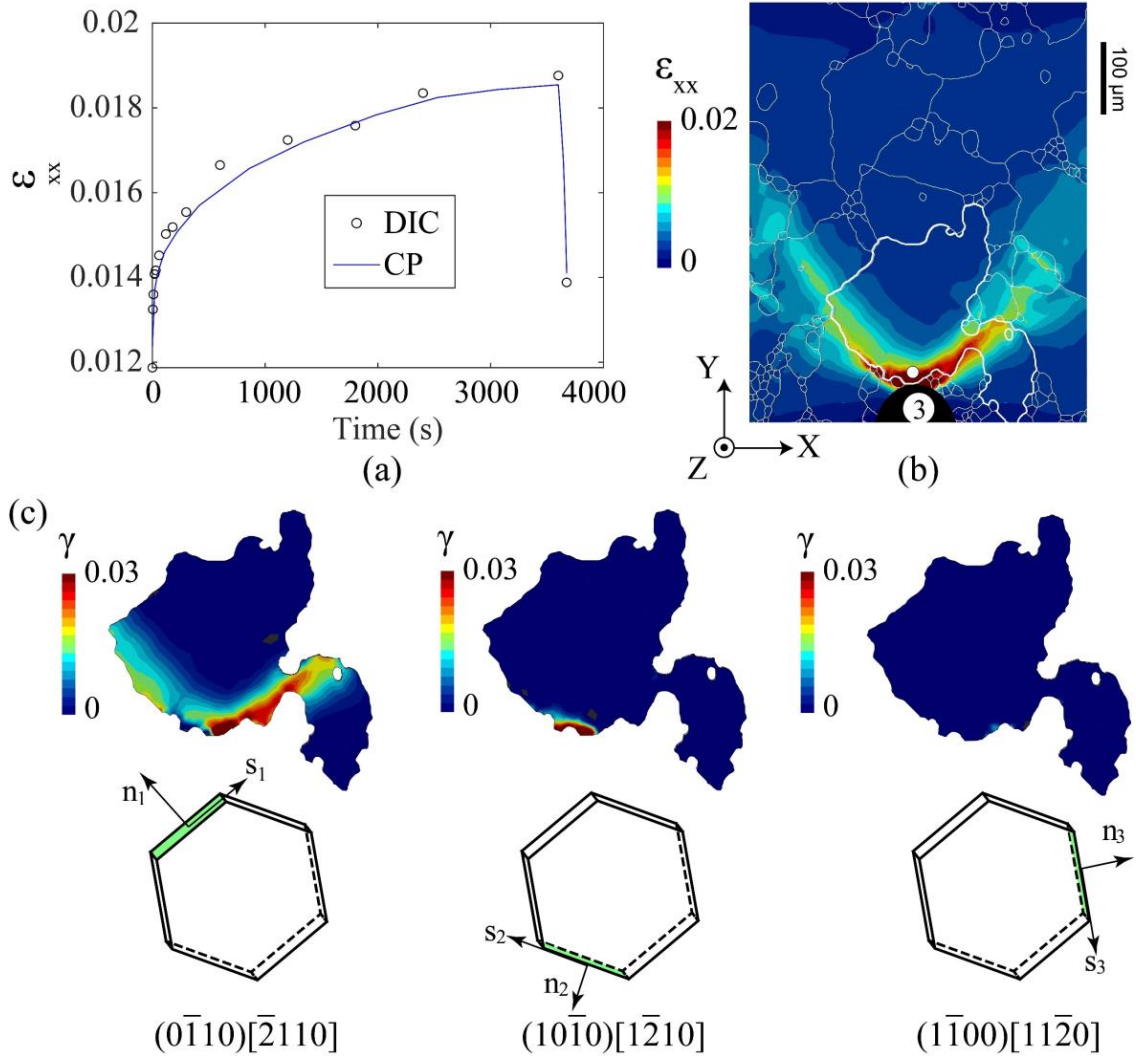

*Figure 9 shows (a) local creep strain $\varepsilon_{xx}$ evolution results of CP and DIC at spot 3; (b) strain $\varepsilon_{xx}$ distribution at the end of stress hold and the spot 3 location with high creep strain at notch tip; (c) local slip analysis inside blocky-α grain near notch.*

At the unloaded stage in Figure 10 (f), the elastic strains are recovered to show the remaining local plastic strain distribution. The remaining plastic strain distribution still shows the 'butterfly wing' pattern which develops along the grain boundaries and reflects in addition the largely prism slip activation. Microstructural impact on the plastic strain distribution is observed at the boundary between grains A and D with misorientation of 11.5°, and at other adjacent grain boundary regions.

Compared with experimental DIC measurements, CP modelling mostly captures the historical strain distribution, including slip localisation band pattern, asymmetry feature development across grain boundaries and major strain concentration near notch tip. The difference in concentration spreading is mainly caused by the roughness of the notch tip with extra grain-level defects and the minor difference in slip band width could be caused by factors including systematic error in DIC measurement with respect to sample tilt angle and current misorientation angle limit of 5°. Decreasing misorientation angle limit results in more precise grain morphologies and 2-3° angle changes in local crystallographic orientation within grains for CP modelling.

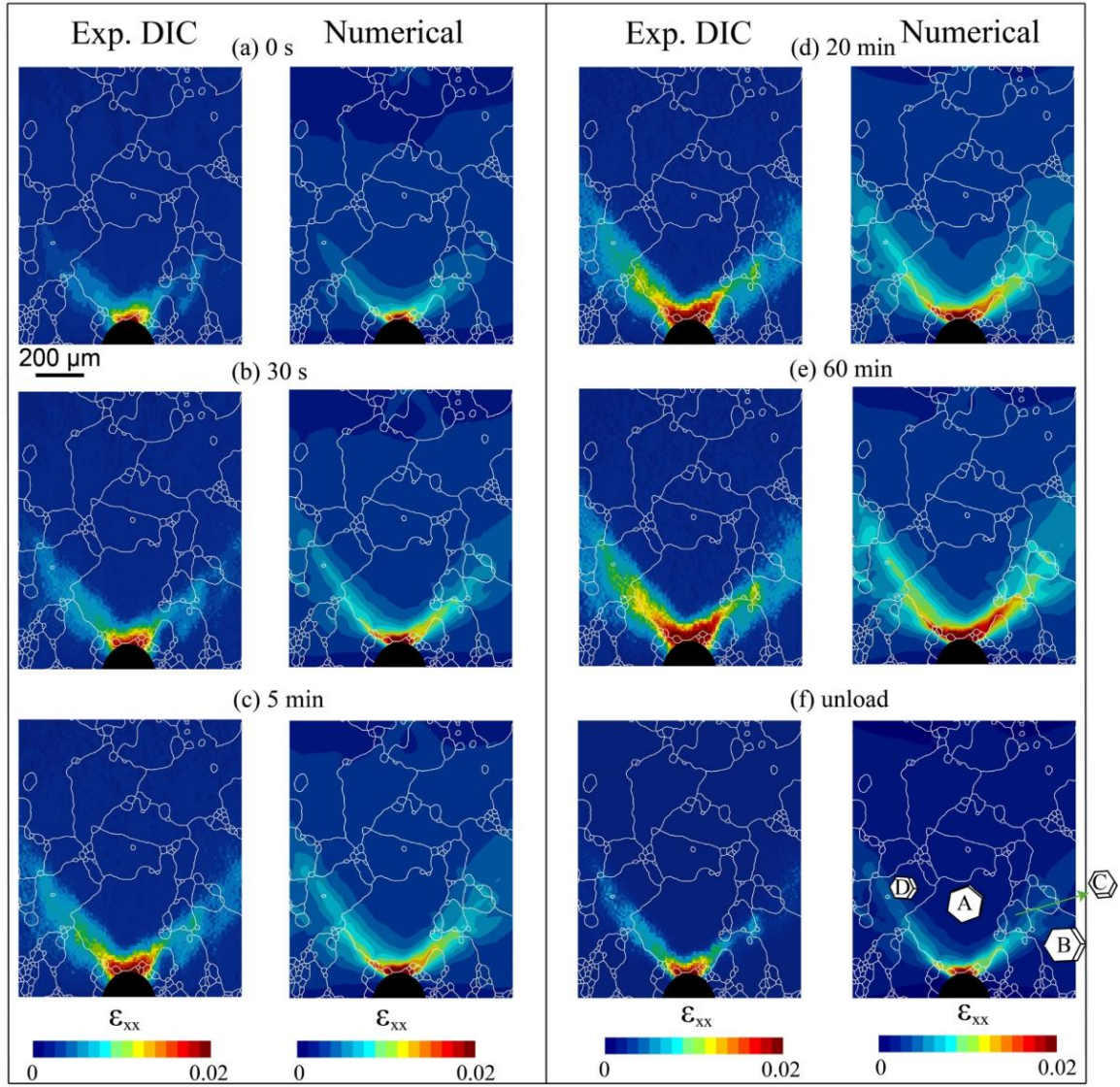

*Figure 10 shows creep responses (strain component $\varepsilon_{xx}$) from experimental DIC measurement and CP modelling at the points during the stress hold time of (a) 0 s, (b) 30 s, (c) 5 min, (d) 20 min, (e) 60 min and (f) at unloaded state.*

### 5.2 Grain-level microscale SRS distribution

In this section, the historical spatial distribution of SRS $m$ and effective plastic strain rate $\dot{\bar{\varepsilon}}^p$ near the left slip-localisation band are shown in Figure 11 (a-h) for different holding time frames of 0 s, 30 s, 120 s and 1800 s. As mentioned in Eq. (7), the SRS $m$ is a variable function based on local plastic strain rate $\dot{\varepsilon}^p$, plastic strain $\varepsilon^p$, local crystallography and temperature. From its analytical form shown in Eq. (12), the local SRS value *m* could be spatially obtained. It is noted that Schmid factor $M^s$, magnitude of Burges vector $b^s$ and CRSS maximum $\tau_c^s$ are determined based on the highest dislocation slip. High SRS region (*m*>0.03) in Figure 11 (a-d) follows the localised slip band region ($\varepsilon_{xx}$>0.01) shown in Figure 10 and the high effective plastic strain rate ($\dot{\bar{\varepsilon}}^p$>2×10$^{-4}$/s) in Figure 11 (e-h), which indicates that high microscale SRS is dominated by grain-level plastic deformation. Misorientation angles are relatively low, ranging from 5.8° to 11.5° among grains shown in this plot, but several grain boundaries still show a clear 'blocking' effect on local SRS distribution shown in Figure 11 (a-c), exhibiting the crystallographic sensitivity for SRS. This agrees with experimental measurement of SRS (Wan, 2020). Heterogenous plastic strain rate $\dot{\varepsilon}^p$ shows consistent critical locations with microscale SRS, which indicates that the crystallographic sensitivity of SRS is the result of different dislocation slip activation across grain boundaries.

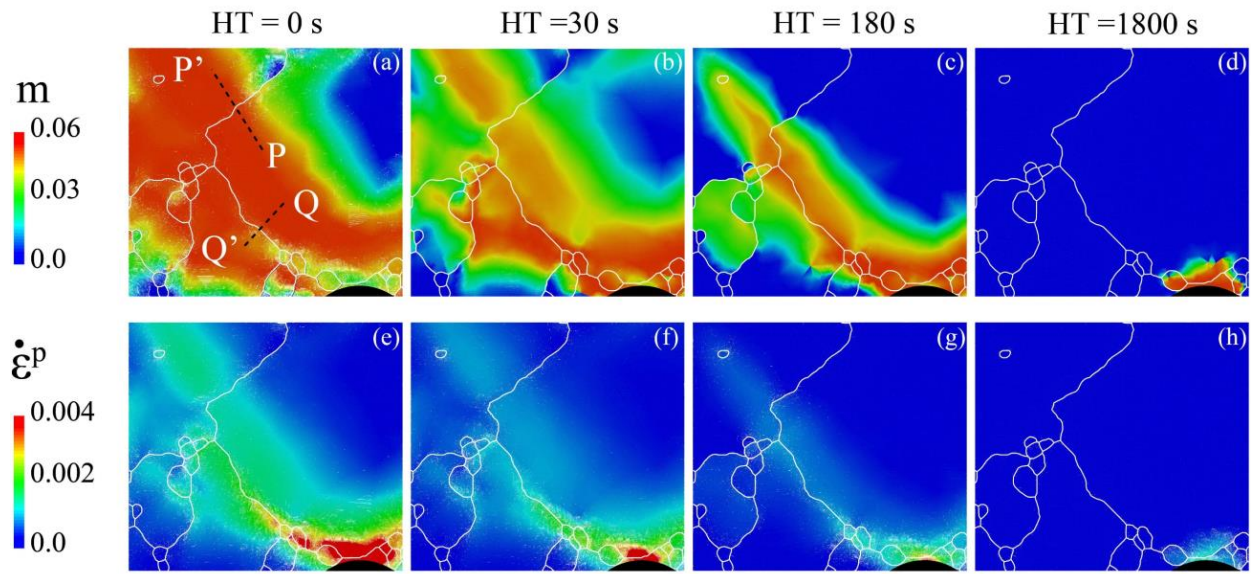

*Figure 11 shows the spatial distribution of (a-d) SRS m and (e-f) effective plastic strain rate $\dot{\bar{\varepsilon}}^p$ at holding time (HT) of 0s, 30s, 120s and 1800s.*

Microscale SRS is the highest at the start of load hold in Figure 11 (a) and its magnitude is decreasing rapidly, which shows the same trend in plastic strain rate $\dot{\varepsilon}^p$ in Figure 11 (e-h), but opposite to strain accumulation inside slip band in Figure 10. To closely investigate the time dependence of SRS *m*, accumulated plastic strain $\varepsilon^p$, and plastic strain rate $\dot{\varepsilon}^p$, their line profiles along path P-P' (Figure 11(a)) are shown in Figure 12 (a-c) at different load hold time (HT) periods, where path P-P' is inside the highest slip localisation region of Figure 10. It is observed that the maximum *m* level is obtained at the start of load hold in Figure 11 (a), which is the time step of lowest $\varepsilon^p$ and highest $\dot{\varepsilon}^p$ in Figure 12 (b) and (c), respectively, due to primary creep occurring along path P-P' where the creep strain increase rate is the highest. A quick drop of SRS is shown for just 30 s hold time from ~0.06 to ~0.03 due to one order of magnitude decrease of creep strain rate from 1.2×10$^{-3}$/s to 2×10$^{-4}$/s. In this time period, local creep strain rises 50%-90% locally in path P-P' in Figure 12 (b). After 120s, the local strain increase tends to stabilise in the slip band. The strain increase rate from 120 s to 1800 s (about 5×10$^{-5}$/s) is negligible compared with that in the first 30 s, reaching steady-state secondary creep stage. This leads to low level of SRS of about 0.002. It demonstrates that SRS is not a material constant but has a strong time-dependence locally at grain level and its evolution is mainly controlled by the plastic strain rate.

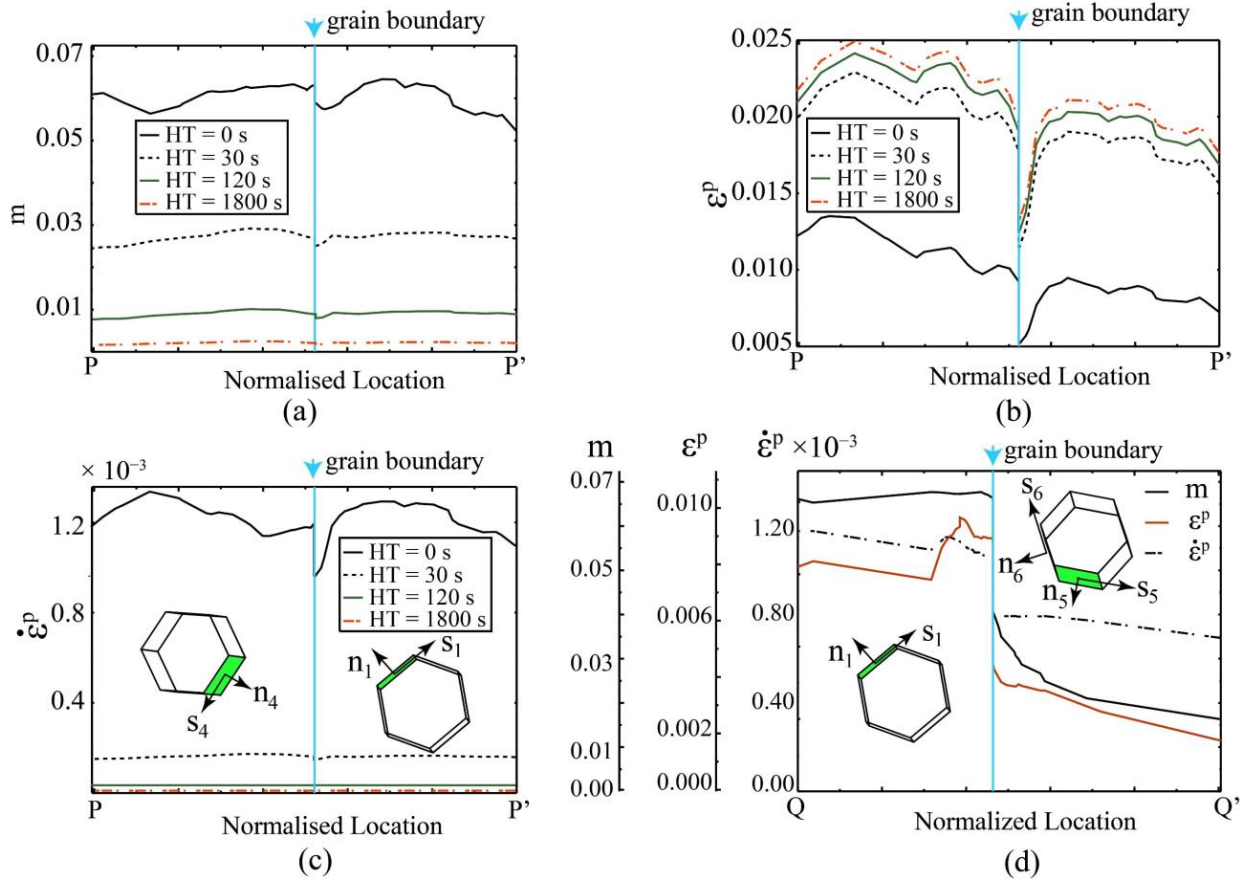

*Figure 12 (a), (b) and (c) show the historical line profiles along path P-P' of m, accumulated plastic strain $\varepsilon^p$, and $\dot{\varepsilon}^p$, respectively. (d) the line profile of path Q-Q' at the start of hold time. Major slip system activations (highest magnitude) are shown in (c) and (d) for local grains.*

The crystallographic sensitivity of SRS across grain boundaries is further analysed. At the grain boundary of path P-P' with misorientation angle of 11.5°, SRS *m*, $\varepsilon^p$, and $\dot{\varepsilon}^p$ all have a sudden drop at the start of load hold, shown in Figure 12 (a-c). But the change of SRS is relatively small, ~5% at grain boundary (b), compared with the significant drop of strain magnitude, reaching ~50% at the same grain boundary (c) and ~20% drop in $\dot{\varepsilon}^p$. Figure 12 (d) shows line profile of *m*, $\varepsilon^p$, and $\dot{\varepsilon}^p$ along path Q-Q' (Figure 11(a)) at the start of load hold, where misorientation angle is similar of 12.3° at the grain boundary of path Q-Q', but the drop of SRS is over 50%, much stronger than that of path P-P', whereas similar dropping percentage are observed for $\varepsilon^p$, which shows more pronounced crystallographic sensitivity in SRS along Q-Q'. It is observed that strain rate gradient difference is relatively small along path P-P' and similar major dislocation slips are activated across grain boundary shown in Figure 12 (c) where angles between slip directions and slip normals within these two activated slip systems are both <15°, indicating favourable slip transfer across grain boundary. Whereas path Q-Q' is perpendicular to the slip band, showing stronger strain rate gradient difference and different major dislocation slip activations across grain boundary in Figure 12 (d) where angles between slip directions and slip normals within activated slip systems across grain boundary are all >50°, exhibiting deviation of slip activation along path Q-Q'. Thus, it indicates SRS is more sensitive to grain crystallography when higher strain rate gradient exists, specifically near slip localisation bands. Besides, it demonstrates that different dislocation slip activation across grain boundaries is the main reason for crystallographic sensitivity in SRS. In addition, it is noted that the crystallographic sensitivity of SRS diminishes when reaching towards secondary steady-state creep shown in Figure 12 (a).

### 5.3 Redistributed stresses after creep

The DIC and CP results show significant microscale creep near the notch for Zircloy-4, which potentially leads to stress redistribution. Stress redistribution mechanism dominated by microscale creep and SRS is crucial for illustrating the local stress state near stress raisers of Zircaloy sample, such as crack nucleation or hydride precipitation sites, during its in-service or storage condition of long stress holding at high temperature. To obtain the grain-level experimental stress state, the stress tensor $\boldsymbol{\sigma}$ at the end of stress hold is computed using

$$\boldsymbol{\sigma} = \mathbf{C} : (\boldsymbol{\varepsilon} - \boldsymbol{\varepsilon}^p) \tag{13}$$

where $\boldsymbol{\varepsilon}$ is the strain tensor at the end of stress hold from DIC measurement and $\boldsymbol{\varepsilon}^p$ is the plastic strain tensor at the unload state. $\mathbf{C}$ is the initial elastic matrices based on the assumption that crystallographic orientation remains the same at local spots. Here, the stress component $\sigma_{xx}$ along principal stress direction is extracted and shown in Figure 13 (a), where the redistributed stress mostly concentrates at grain boundaries near notch tip.

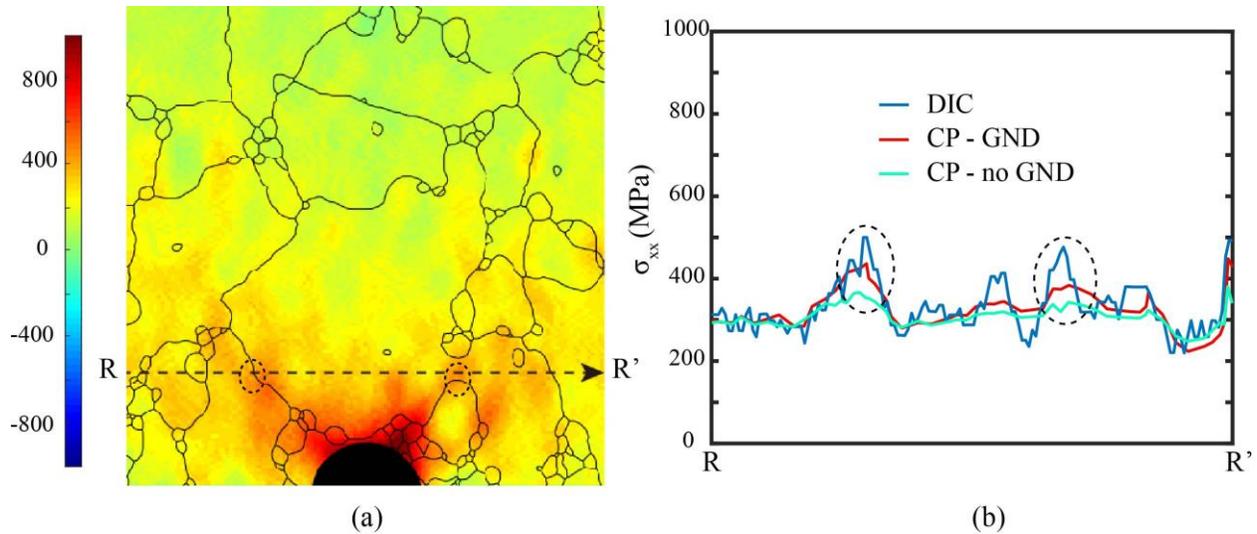

*Figure 13 (a) experimental stress $\sigma_{xx}$ state of ROI near notch tip; (b) $\sigma_{xx}$ comparison along path R-R' between CP and DIC results.*

The CP results shows a good agreement with the DIC results, which demonstrates that thermally activated dislocation slip in flow rule from Eq. (5) is sufficient to capture the rate-dependent and microstructure-sensitive stress responses at crystal level. The dotted lines in Figure 13 (a) point out two positions near grain boundaries with misorientation of 12.3° and 9.5° showing a raise in $\sigma_{xx}$ levels of both DIC and CP model considering GND effects. GND hardening effect on CRSS in Eq. (6) is crucial to accurately describe the local strain state where elastic strain is redistributed to accommodate local lattice curvature change (Arsenlis and Parks, 1999) which occurs during geometrical dislocation pinning near grain boundaries from intense grain interaction. The CP modelling without GND effect leads to much higher deviation of local stress profile near corresponding dotted region in Figure 13 (b) where GND is needed to accommodate high curvature near grain boundaries in Figure 13 (a).

## 5.4 Temperature sensitive texture effect

Temperature-sensitive texture effect on polycrystalline responses is investigated in this section. The heterogeneous stress $\sigma_{yy}$ distribution at 623 K is shown in Figure 14 (a), which exhibits different stress concentration locations from Figure 8 (a) loaded along TD at same temperature. Employing same extracted crystal properties in Table 1, polycrystalline RVE responses in Figure 6 (a) are extracted using CP model loaded along RD. Polycrystalline stress-strain curves from CP agree well with experimental strain-strain curves at both temperatures along RD in Figure 14 (b). This demonstrates the capability of extracted single-crystal properties in capturing both temperature and texture sensitivity of polycrystal when applying faithful statistical representations of the textures and grain geometries.

Temperature-sensitive yielding stress $\sigma_{0.2}$ at offset strain of 0.2%, strain hardening rate $d\sigma/d\varepsilon$ and their dropping magnitudes $\Delta\sigma_{0.2}|_{\Delta T}$ and $\Delta(d\sigma/d\varepsilon)|_{\Delta T}$ with respect to temperature increase $\Delta T = 330$ K are obtained in Table 2 for both loading directions. Yielding stresses $\sigma_{0.2}$ of TD direction show higher magnitudes at both temperatures. On contrary, hardening rates $d\sigma/d\varepsilon$ are higher along RD at both temperatures. Strain hardening rate along TD at 623 K is much lower than the other three conditions, which leads to stress saturation quickly at strain of 0.01 in Figure 14 (b). In addition, the dropping magnitude of both $\sigma_{0.2}$ and $d\sigma/d\varepsilon$ are higher in TD when temperature increases.

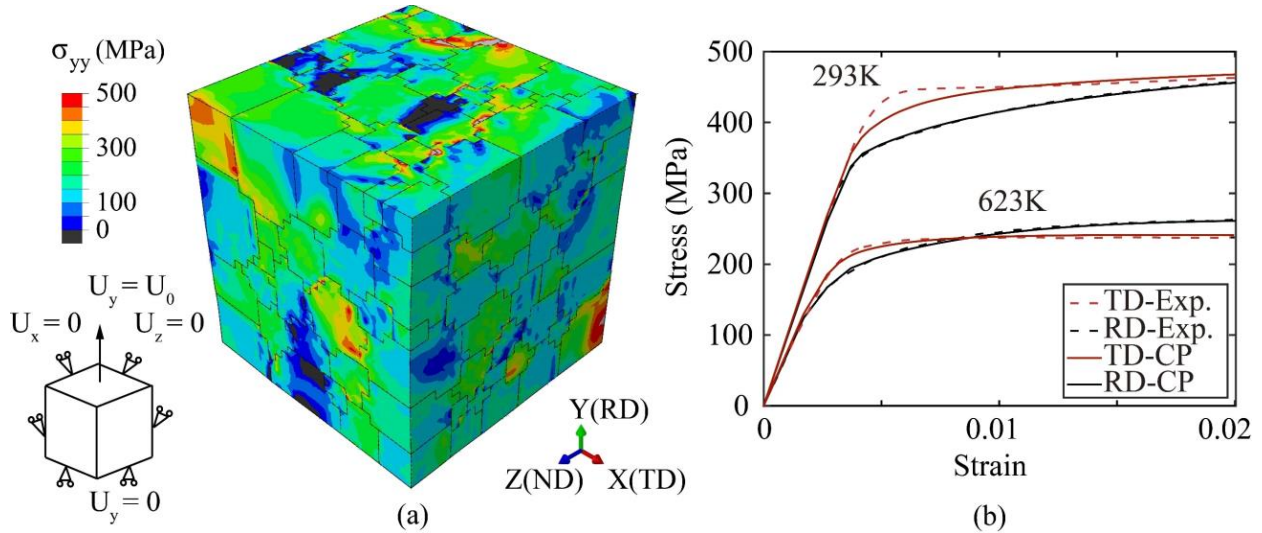

*Figure 14 (a) stress $\sigma_{yy}$ state of RVE at the end of strain 0.02 tensile at 623 K along RD direction. (b) stress-strain comparison between tensile experiments and CP results loaded along TD and RD directions (a) and (b) are both at 0.01/s strain rate.*

*Table 2 Temperature-sensitive macroscale properties*

|  | RD 293 K | RD 623 K | TD 293 K | TD 623K |
|---|---|---|---|---|
| $\sigma_{0.2}$ (MPa) | 367 | 190 | 413 | 210 |
| $d\sigma/d\varepsilon$ (MPa) | 83 | 72 | 48 | 19 |
| $\Delta\sigma_{0.2}|_{\Delta T}$ (MPa) | 177 | | 203 | |
| $\Delta(d\sigma/d\varepsilon)|_{\Delta T}$ (MPa) | 11 | | 29 | |

To better understand the polycrystalline response, crystal-level strain accumulations are obtained in all types of HCP slip systems in Zircaloy-4, including prism, basal, pyramidal <a> and pyramidal <c+a>, for the four loading conditions, shown in Figure 15 (a-d). The total shear strain accumulation for one system type is time-integrated as $\gamma_{tot} = \int_0^t \sum_{s=1}^r \dot{\gamma}^s \, dt$, where $r$ is the total slip system number of this type and $\dot{\gamma}^s$ is computed in Eq. (5). Total dislocation density $\rho_{tot}$ is extracted for all the grains inside polycrystal including $\rho_{GND}$ and $\rho_{SSD}$, shown in Figure 15 (e).

Figure 15 (a, c) show the sufficient pyramidal <c+a> slip evolution when loaded along TD whereas no <c+a> slip is observed along RD in Figure 15 (b, d). This is caused by the relative angle between c axes of grains and loading direction. Shown in Figure 6 (b), more c axes of grains align near TD direction, but mostly perpendicular to the RD, which induces favourable pyramidal <c+a> slip loaded under TD load when local grains have c axes nearly parallel to loading direction. Pyramidal <c+a> slip has about two times higher CRSS than that of <a> slip systems (Gong et al., 2015), which contribute to the relatively higher polycrystalline yielding stress in TD load from Table 2. Besides, the pyramidal <c+a> slip is more sensitive to temperature, meaning higher absolute CRSS magnitude decreasing when temperature increases. This leads to higher yielding stress dropping along TD in Table 2 compared with that of RD.

Total shear strain accumulations $\gamma_{tot}$ for the four loading conditions all show high prismatic slip activation. The shear strain accumulation along TD is inclined to be decomposed into pyramidal <a> and <c+a> slip systems, which lowers the prismatic slip accumulation level in this direction. More types of slip system activation are observed at higher temperature in both loading directions, i.e., pyramidal <a> rising in TD and basal <a> rising in RD when temperature increases.

Figure 15 (e) shows the total dislocation density $\rho_{tot}$ evolution in four loading conditions. Though a relatively high $\gamma_s$ in Table 1 is extracted for $\rho_{SSD}$ evolution, compared with small hardening effect of $\rho_{SSD}$ in other single crystal HCP, e.g. titanium (Zhang et al., 2016a, 2016b), it is noted that hardening effect of $\rho_{GND}$ is dominant for CRSS increase during dislocation slip accumulation. In Figure 15 (e), the dislocation density level is much higher along RD compared with that of TD, which explains the higher strain hardening rate $d\sigma/d\varepsilon$ in RD from Table 2 since more pinning dislocations are achieved due to intense prismatic slip interactions in this direction. The total dislocation density is especially low in TD at 623K, which is consistent with the low hardening rate at this condition in Table 2. $\rho_{tot}$ in TD decreases to less than half of its magnitude when temperature increase, which leads to the higher dropping of $\Delta(d\sigma/d\varepsilon)|_{\Delta T}$ in TD from Table 2.

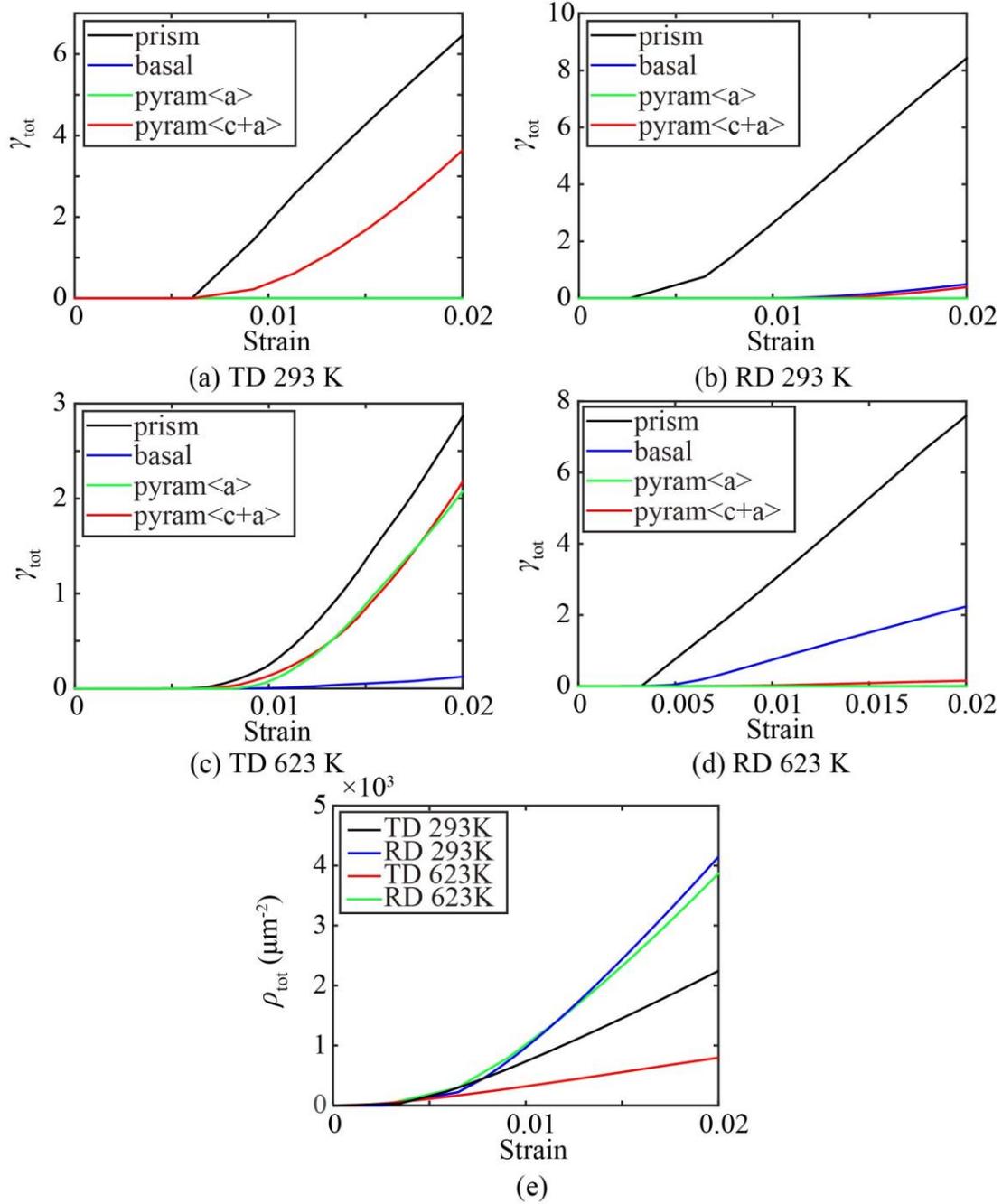

*Figure 15 shows the total shear strain accumulations $\gamma_{tot}$ inside every slip system for loading conditions of (a) TD and 293 K, (b) RD and 293 K, (c) TD and 623 K, and (d) RD and 623 K. (e) Total dislocation density $\rho_{tot}$ accumulation for the four loading conditions.*

## 6. Discussion

To discuss SRS in different scales, the microscale SRS is upscaled to polycrystalline SRS and lattice-scale SRS in separate multi-hold tests. Then the uncertainties of dislocation-based crystal plasticity method in capturing multi-scale rate and temperature sensitivities are discussed.

## 6.1 Upscaling SRS to macroscale

In this section, the spatially varying microscale SRS is upscaled to macroscale using an independent multi-hold test. Same calibrated single-crystal properties in Table 1 is employed during the upscaling and the grain-representative polycrystal is reconstructed based on the grain size and crystallographic information from experiments (Kumar et al., 2018) and the macroscale polycrystalline responses are extracted. Based on geometrical grain statistics, the reconstructed polycrystals with 300 grains is shown in Figure 16 (a). Regarding crystallographic info, the grain orientations are imported using experimental data of textured Zircaloy-4 with the same Kearns factors along three orthogonal directions, i.e., $f_x = 0.085$, $f_y = 0.185$, $f_z = 0.73$ (Kumar et al., 2018).

CP study is carried out for reported multi-strain-hold tests conducted at room temperature with strain rate of $1\times10^{-3}$/s along the TD (Kumar et al., 2018). The multi-strain hold allows for easily obtaining the macroscale value of $m$ at several different strain magnitudes from one single sample, eliminating potential microstructural effects among different samples. The polycrystalline stress-time curve is obtained by averaging historical stress response of component $\sigma_{xx}$ along TD load direction among all integration points in CP model, which agrees well with the experimental data in Figure 16 (b), especially the stress drop during each strain hold is well captured.

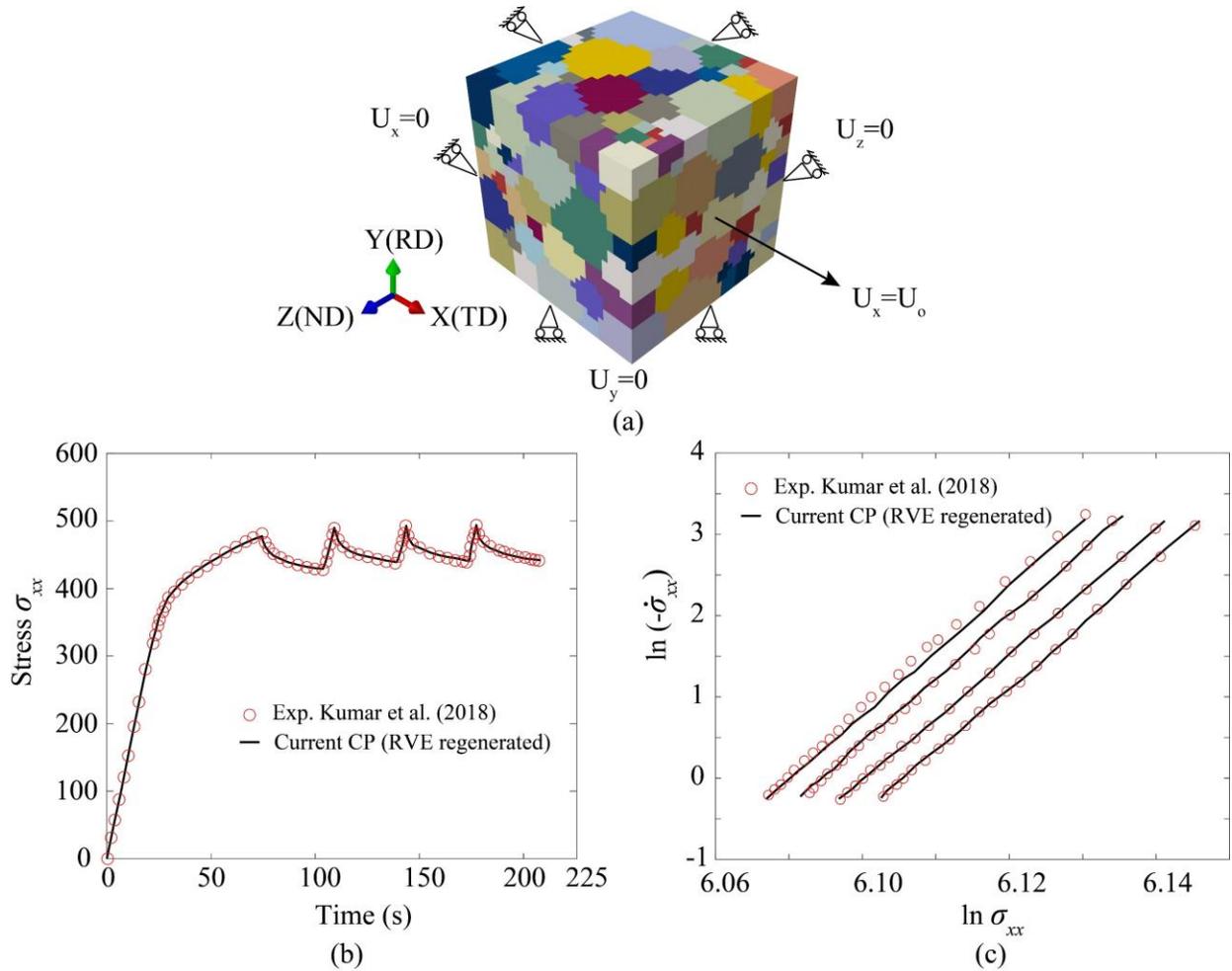

*Figure 16 shows the present model capture of experimentally measured macroscale strain rate sensitivity in an independent study of textured zircloy-4 sample. (a) Regenerated polycrystalline structure for reference paper (Kumar et al., 2018); (b) σ$_{xx}$~t comparison during the strain-controlled load; (c) $ln(-\dot{\sigma}_{xx}) \sim ln(\sigma_{xx})$ comparison during the four strain hold periods.*

To further investigate the time-dependent stress evolution, the relationship between stress dropping rate and stress level is plotted in logarithmic form ($ln(-\dot{\sigma}_{xx})$ versus $ln\,\sigma_{xx}$) at four strain hold periods of 0.075, 0.0775, 0.080 and 0.0825. The predicted curves show close agreements with experimental data (Kumar et al., 2018) and the macroscale SRS is quantified as 0.03 at strain rate of $1\times10^{-3}$ /s, which is compatible with microscale SRS shown in Figure 12 (a) and (c). This demonstrates that upscaling microscale SRS leads to the consistent macroscale SRS of polycrystal. Besides, when providing the grain statistical information including grain size and crystallography, the macroscale SRS of Zircaloy-4 could be captured accurately using single-crystal properties extracted from Section 4.

### 6.2 Orientation-sensitive lattice SRS in XRD for Zircaloy-2

In this section, the 'virtual' in-situ XRD monotonic polycrystal modelling is developed and the lattice SRS $m_{\{hkil\}}$ is analysed using CP compared with independent experimental measurements using in-situ XRD measurements (Skippon et al., 2019). A textured Zircaloy-2 sample was subjected to multi-strain-hold tests along both RD and ND directions (Skippon et al., 2019), which provided experimental uniaxial average stress-strain data, and these were used to determine crystal slip properties for this material. The crystal properties for Zircaloy-4 from Section 4 are unchanged with the exception that the slip system activation energy Δ$F$ and corresponding activation volume Δ$V$ are tuned to ensure correct capture of the Zr-2 alloy macroscale SRS in multi-hold test. Figure 17 (a) shows the RVE microstructure assigned the microstructural statistics as observed for the Zircaloy-2 study of Skippon et al. (2019) for strain-controlled loading along the RD direction.

It is noted that SRS would normally be anticipated to be determined from *plastic* strain rate data, but the experimental methodology (Skippon et al., 2019) utilises strain hold tests macroscopically, it is reasonable to assume the elastic and plastic strains are equal in magnitude and opposite in sign. How accurate an assumption this is at the length scale of intragranular slip plane deformation is potentially arguable but in mitigation, given lattice strain measurements were averaged over grain clusters with similarly oriented planes within the beam zone.

A brief summary of the XRD methodology is included here to illustrate the numerical procedure for computing lattice SRSs from the CP model. The lattice plane spacing *d* is computed based on the crystal geometry,

$$d_{\{hkil\}} = \frac{a}{\sqrt{\frac{4}{3}(h^2 + hk + k^2) + \frac{a^2 l^2}{c^2}}} \quad (14)$$

where *a* and *c* are the geometrical HCP crystal parameters and *h, k, i, l* are the lattice plane indices. Here, the spacing *d* for the diffraction peaks observed in Zircaloy-2 is between 1.5 and 3.0 Å and the experimental X-ray energy ($E$ = 70 keV) and corresponding wavelength ($\lambda$ = 0.177 Å) from Cochrane et al. (2019) are adopted to compute the incident angles $\phi$, so as to determine the grains of interest with preferable lattice planes that trigger the diffraction peaks, given by

$$\phi = a\sin\left(\frac{(n)\lambda}{2d_{\{hkil\}}}\right) \quad (15)$$

where *n* is a positive integer. The 2$\phi$ angle between the incident and reflection directions is shown in Figure 17 (a). The model polycrystal texture is reconstructed to reproduce that for Zircaloy-2 in Skippon et al. (2019) shown in Figure 17 (b) and the multiple strain hold loading is also adopted with the two experimental strain rates of $1.0\times10^{-4}$ s$^{-1}$ and $1.0\times10^{-5}$ s$^{-1}$ shown in Figure 17 (c). The duration (480 s) of each individual hold time is assigned the same value as the experimental strain rate tests. Two different loading directions along RD and ND directions were investigated in the experiment to study the texture effect on the mechanical responses. In the CP model used here, the slip rule activation volume $\Delta V$ and activation energy $\Delta F$ properties are chosen to be $\Delta V = 16.03$ $b^3$ and $\Delta F = 0.25$ eV to best represent the macroscale stress-strain response and SRS of the Zircaloy-2 alloy investigated in Skippon et al. (2019), which results in similar polycrystalline stress-strain curves of multiple displacement holds compared with experimental data shown in Figure 17 (d) and (e) for ND and RD loading directions, respectively, especially the stress relaxations during each displacement hold periods.

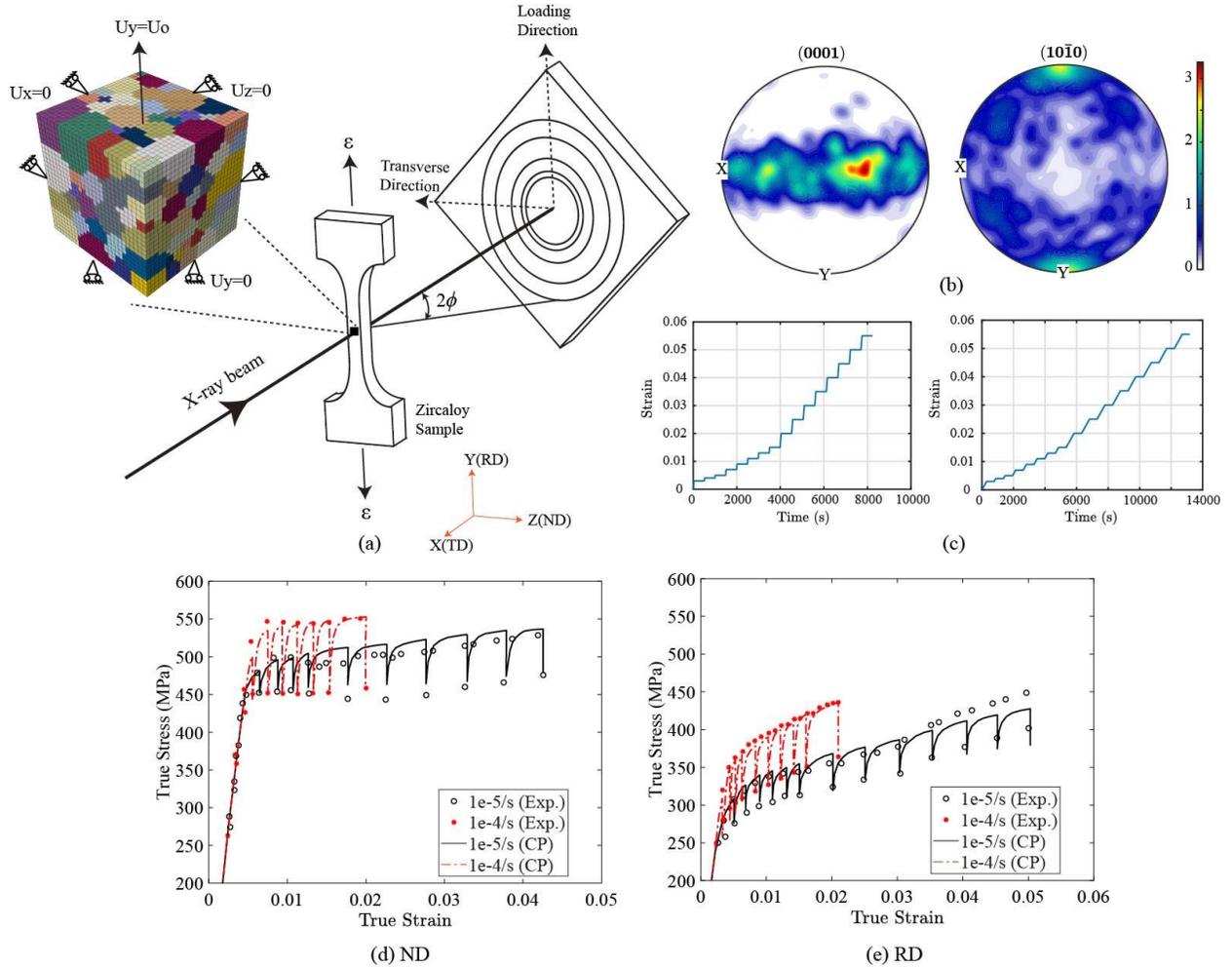

*Figure 17 (a) the test set-up in Skippon et al. (2019) with in-situ XRD measurement, (b) the model reconstructed pole figures showing close correspondence to the Zircaloy-2 texture from Skippon et al.,*

*(2019), (c) the applied multi-hold strain control of the sample test under strain rates of $1.0\times10^{-4}$ $s^{-1}$ (left) and $1.0\times10^{-5}$ $s^{-1}$ (right), and in (d) and (e) the stress-strain responses extracted from CP and experiments (Skippon et al., 2019) along the ND and RD directions, respectively.*

In experimental measurement, a homogeneous uniaxial stress state is assumed within any given grain, such that the individual lattice plane SRS is computed from the lattice space change measurements during the test (Skippon et al., 2019), i.e.,

$$m_{\{hkil\}} = \frac{d \ln \varepsilon^e_{\{hkil\}}}{d \ln(-\dot{\varepsilon}^e_{\{hkil\}})} \tag{16}$$

where $\varepsilon^e_{\{hkil\}} = (d-d_0)/d_0$ and $d_0$ is the initial lattice spacing of {*hkil*} lattice plane. The SRS is defined in terms of *elastic* strain rates as opposed to the plastic strain rate. To extract the equivalent lattice rate sensitivity $m_{\{hkil\}}$ with same assumption of homogeneous uniaxial stress state in CP model, the numerical elastic strain component $\varepsilon^e_{\{hkil\}}$ normal to a given plane is determined from full elastic strain tensor $\boldsymbol{\varepsilon}^e$ with knowledge of the normal direction $\mathbf{n}_{\{hkil\}}$ of the {*hkil*} lattice plane, that is

$$\varepsilon^e_{\{hkil\}} = (\boldsymbol{\varepsilon}^e \, \mathbf{n}_{\{hkil\}}) \cdot \mathbf{n}_{\{hkil\}} \tag{17}$$

For each diffraction peak corresponding to a crystallographic plane {*hkil*}, clusters of grains are extracted satisfying relationship between incident angles $\phi$ and lattice spacing $d_{\{hkil\}}$ in Eq. (19).

To compute the elastic strain of {*hkil*} crystallographic planes for each grain, misorientation range of 15° is chosen for the specific plane following their methodology (Cochrane et al., 2019). Hence the grain clusters for every crystallographic-plane-associated diffraction peak are determined for both Y and Z-axis loading. Then the volume-averaged crystallographic plane elastic strain component is determined

$$\varepsilon^e_{\{hkil\}} = \sum_{k=1}^{n} \frac{\left(\varepsilon^e_{\{hkil\}}\right)_k}{V_k} \tag{18}$$

where $\left(\varepsilon^e_{\{hkil\}}\right)_k$ is the elastic strain component in the $k^{th}$ grain of the grain cluster for {*hkil*} plane and $V_k$ is the volume of this grain. Here, six diffraction peaks with corresponding crystallographic planes $\{10\bar{1}0\}$, $\{0002\}$, $\{10\bar{1}1\}$, $\{10\bar{1}2\}$, $\{11\bar{2}0\}$, $\{10\bar{1}3\}$ are obtained during the multi-hold tests.

Figure 18 shows the CP predicted lattice SRSs in the six crystallographic planes chosen in the referenced paper and clear texture dependence of rate sensitivity is shown by comparing Z-direction loading in (a,b) and Y-direction loading in (c,d), corresponding to ND and RD directions in Skippon et al. (2019). The ND response shows an initial rapid increase in SRS followed at a strain of ~0.01 by a slow but progressive decrease in SRS for all planes. RD loading response is rather different and shows the same initial rapid increase in SRS but followed at a strain of ~0.01 by a stabilization of SRS. In later stages of the multi-holding test, the predicted lattice SRS $m_{\{hkil\}}$ is typically higher for RD loading than for ND. These overall trends from CP are consistent with the experimental observations (Skippon et al., 2019).

The lattice rate sensitivity $m_{\{hkil\}}$ for the $\{10\bar{1}2\}$ crystallographic plane is higher along RD than that for basal $\{0002\}$ and prismatic $\{10\bar{1}0\}$, $\{11\bar{2}0\}$ planes, especially at lower strain rates in Figure 18 (d). The high $m_{\{hkil\}}$ for $\{10\bar{1}2\}$ is observed along ND, but the spreading among all lattice plane SRSs is smaller. Prismatic planes tend to show lower lattice SRSs than that in basal and pyramidal planes in both load directions. The higher SRSs for the pyramidal <c+a> planes and lower SRSs for prismatic planes agree well with the experimental observations. ND load shows higher SRS for higher applied strain rate in the predictions similar with self-consistent modelling in Skippon et al. (2019), which might be caused by low diffraction peak intensity for $\{0002\}$ plane.

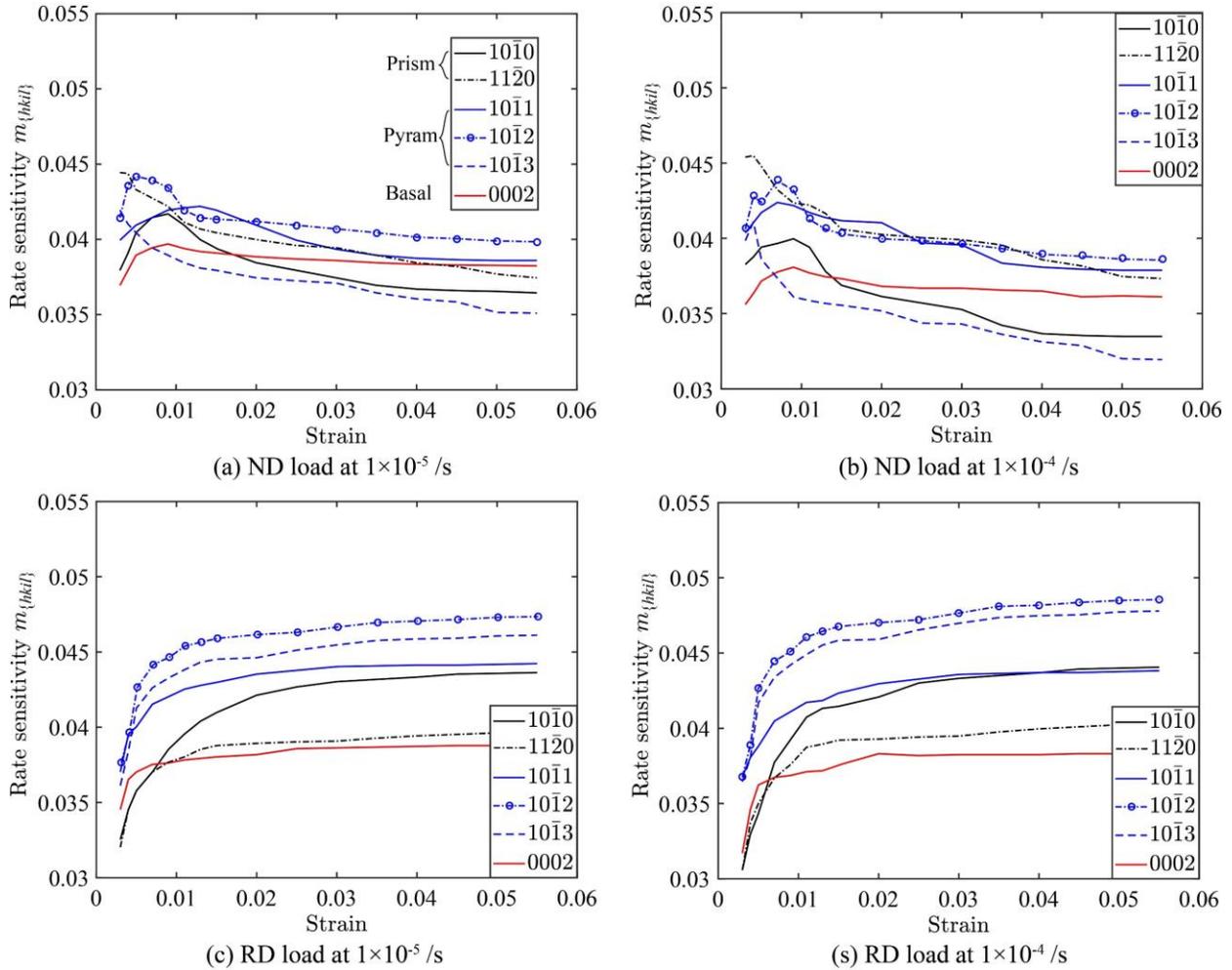

*Figure 18 CP predicted crystallographic-plane-based lattice SRS versus true strain for Zircaloy-2 for loading directions along (a, b) ND and (c, d) RD at strain rate of (a,c) $1.0\times10^{-5}$ $s^{-1}$ and (b,d) $1.0\times10^{-4}$ $s^{-1}$ for the crystal planes shown where the black, red and blue colours represent the prismatic, basal and pyramidal systems respectively. (See Skippon et al. (2019) for independent corresponding XRD-measured SRSs.)*

In the methodology for calculating lattice SRS $m_{\{hkil\}}$ in Skippon et al. (2019), it is necessary to assume that a uniaxial stress state exists in each grain in order to calculate the corresponding elastic strain and strain rate, and hence SRS. This simplification is not needed in the more general CP approach, but still CP provides good agreement with the experimentally determined $m_{\{hkil\}}$ based on the assumption. The chemical compositions of Zircaloy-2 and Zircaloy-4 are quite similar with 0.05% percentage difference of Fe and Ni, but this small chemical composition differences seem to lead to different lattice SRSs between Zircaloy-4 and Zircaloy-2. The CP predicted polycrystalline SRS of Zircaloy-4 at strain rate of $1.0\times10^{-4}$ s$^{-1}$ is about 0.015, is much lower than that of Zircaloy-2 calculated from CP, which are consistent with experimental measurements of Zircaloy-4 (0.07-0.016 reported in (Lee et al., 2001)) and Zircaloy-2 (~0.03 in (Skippon et al., 2019)) at a strain rate of $1.0\times10^{-4}$ s$^{-1}$ at room temperature.

## 6.3 Uncertainties in capturing multi-scale SRS and temperature sensitivity

This section discusses the uncertainties of current method in capturing SRS and temperature sensitivity across scales. The CP method has been demonstrated to reveal the microscale strain rate and temperature sensitivity based on dislocation activation, slip and pinning mechanism. The grain-level microscale SRS is linked to local crystallography, dislocation slip rate and accumulation, which is accounted in individual slip system. Dislocation pinning in CP model is based on statistical dislocation density, however, individual dislocation nucleation, dislocation mobility and thermally-activated dislocation gliding have significant impact on inner-grain strain rate sensitivity for HCP crystals (Zheng et al., 2016). Applying discrete dislocation plasticity (DDP) method allows one to study the effect of thermally activated dislocation motion on dislocation-level rate sensitivity and a closer look in trans-granular SRS change across grain boundaries inside Zircaloy-4 under in-service thermo-mechanical cycles.

The intrinsic anisotropy of rate sensitivity in each slip system is significant for HCP crystals, e.g. for titanium alloys (Jun et al., 2016b; Zhang et al., 2016a). In this study, strain rate sensitivities are extracted based a highly textured blocky-$\alpha$ Zircaloy-4 with dominant prismatic slip activation shown in Figure 9. The micro-scale SRS based on activation volume $\Delta V$ and activation energy $\Delta F$ that controls slip-system-level rate sensitivity has not been studied fully for other slip systems, including basal, pyramidal <a> and pyramidal <c+a>, which requires future micro-pillar testing or cantilever testing showing individual slip activation in micro-scale samples. Besides, the pinning of GND and SSD could leads to instantaneous variation of dislocation mean free path $l=1/\sqrt{\rho_{tot}}$, which in turn changes crystal-level activation volume ($\Delta V=\gamma_0 b^2/\sqrt{\rho_{tot}}$) and inner-grain SRSs (Dunne et al., 2007). The upscaling of SRS from micro- to macro-scale was argued to be 'microscopic stress' in wider range of strain rates from quasi-static $10^{-3}$/s to dynamic $10^5$/s for zirconium (Zecevic et al., 2016). Broaden current CP method to a wider strain rate range for Zircaloy-4, especially the dynamic range, is valuable for understand multi-scale SRS based on thermally activated dislocation slip and dynamic slip bands.

The temperature range studied is relatively low compared to extreme in-service thermal conditions for nuclear reactors. Crystallography and texture-dependent creep behaviour at higher temperature range is beneficial for understanding dislocation-scale mechanism, potentially changing creep mechanism from dislocation glide to dislocation climb or even diffusion creep. In addition, during thermal annealing process, grain-level residual stresses and initial dislocation densities distribution are important to incorporate for characterising microscale SRS since initial pinning dislocations and residual lattice curvature could potentially affect the dislocation slip during practical conditions of Zircaloy-4.

## 7. Conclusion

Intrinsic slip system strain rate sensitivity properties have been extracted for Zircaloy-4 and polycrystal modelling is shown to capture experimentally obtained DIC-measured creep strains, polycrystal stress relaxation, and rate-dependent tensile stress responses and their texture dependence. Room temperature creep bend testing shows significant microscale creep strain development near a notch, indicating high microscale rate sensitivity, different from experimental measured polycrystalline SRSs of < 0.1 up to 700 K in nuclear reactor applications (Lee et al., 2001). This means that even at room temperature for stress hold periods in reactor plant, cracks or notches in the cladding will generate local creep, driving stress state changes and hydrogen uptake which is important for the well-known delayed hydride cracking (Shi et al., 1995).

Microscale SRSs are highly dependent on local plastic strain rate and the crystallographic sensitivity of SRS is resulted from different dislocation slip activations across grain boundaries. Redistributed stress originated from microstructural creep is captured well by CP method, crucial for future investigation of local stress state near stress raisers during in-service or storage long stress hold. The temperature-sensitive texture effect is captured by CP, due to slip-system-based slip accumulation and dislocation pileups. The high CRSS value and temperature sensitivity of pyramidal <$c+a$> slip activation leads to the texture dependency while c-axis pointing towards the loading direction. Multi-scale SRSs could be captured only if a microstructure-faithful model is reconstructed from statistical information of grain morphology and crystallography combined with rate-sensitive single-crystal properties, which demonstrate CP model self-consistency of SRS across different scales.

## Acknowledgements

YL would like to thank Dr Yi Guo for helpful discussions on XRD. The financial support provided by UKRI (EP/S01702X) is also gratefully acknowledged. MW, CMD and SEC acknowledge financial support from Rolls-Royce plc and the EPSRC Centre for Doctoral Training in Nuclear Energy grant no. EP/L015900/1.

## References:

Abdolvand, H., 2019. Progressive modelling and experimentation of hydrogen diffusion and precipitation in anisotropic polycrystals. Int. J. Plast. 116, 39–61. https://doi.org/10.1016/j.ijplas.2018.12.005

Abdolvand, H., Wright, J., Wilkinson, A.J., 2018. Strong grain neighbour effects in polycrystals. Nat. Commun. 9, 1–11. https://doi.org/10.1038/s41467-017-02213-9

Adamson, R.B., 1977. Irradiation Growth of Zircaloy, in: Lowe, A.L., Parry, G.W. (Eds.), Zirconium in the Nuclear Industry. ASTM International, West Conshohocken, PA, pp. 326–343. https://doi.org/10.1520/STP35579S

Arsenlis, A., Parks, D.M., 1999. Crystallographic aspects of geometrically-necessary and statistically-stored dislocation density. Acta Mater. 47, 1597–1611. https://doi.org/10.1016/S1359-6454(99)00020-8

Barrett, T.J., Savage, D.J., Ardeljan, M., Knezevic, M., 2018. An automated procedure for geometry creation and finite element mesh generation: Application to explicit grain structure models and machining distortion. Comput. Mater. Sci. 141, 269–281. https://doi.org/10.1016/j.commatsci.2017.09.048

Chan, K.S., 2013. An assessment of delayed hydride cracking in zirconium alloy cladding tubes under stress transients. Int. Mater. Rev. 58, 349–373. https://doi.org/10.1179/1743280412Y.0000000013

Chen, B., 2018. Understanding Microstructurally-Sensitive Fatigue Crack Nucleation in Superalloys 168.

Chen, B., Jiang, J., Dunne, F.P.E., 2017. Microstructurally-sensitive fatigue crack nucleation in Ni-based single and oligo crystals. J. Mech. Phys. Solids 106, 15–33. https://doi.org/10.1016/j.jmps.2017.05.012

Cheng, J., Shahba, A., Ghosh, S., 2016. Stabilized tetrahedral elements for crystal plasticity finite element analysis overcoming volumetric locking. Comput. Mech. 57, 733–753. https://doi.org/10.1007/s00466-016-1258-2

Choi, Y.S., Groeber, M.A., Turner, T.J., Dimiduk, D.M., Woodward, C., Uchic, M.D., Parthasarathy, T.A., 2012. A crystal-plasticity FEM study on effects of simplified grain representation and mesh types on mesoscopic plasticity heterogeneities. Mater. Sci. Eng. A 553, 37–44. https://doi.org/10.1016/j.msea.2012.05.089

Cochrane, C., Skippon, T., Daymond, M.R., 2019. Effect of rate on the deformation properties of metastable β in a high Sn content zirconium alloy. Int. J. Plast. 119, 102–122. https://doi.org/10.1016/j.ijplas.2019.02.019

Cottrell, A.H., Dexter, D.L., 1954. Dislocations and Plastic Flow in Crystals. Am. J. Phys. 22, 242–243. https://doi.org/10.1119/1.1933704

Cuniberti, A., Picasso, A., 2001. Activation volume measurement techniques: application to Zircaloy-4. Phys. Status Solidi Appl. Res. 183, 373–379. https://doi.org/10.1002/1521-396X(200102)183:2<373::AID-PSSA373>3.0.CO;2-N

Deka, D., Joseph, D.S., Ghosh, S., J., M.M., Mills, M.J., 2006. Crystal Plasticity Modeling of Deformation and Creep in Polycrystalline Ti-6242. Metall. Mater. Trans. a 37A, 1371–1388. https://doi.org/10.1007/s11661-006-0082-2

Derep, J.L., Ibrahim, S., Rouby, R., Fantozzi, G., 1980. Deformation behaviour of zircaloy-4 between 77 and 900 K. Acta Metall. 28, 607–619. https://doi.org/10.1016/0001-6160(80)90127-3

Dunne, F.P.E., Rugg, D., Walker, A., 2007. Lengthscale-dependent, elastically anisotropic, physically-based hcp crystal plasticity: Application to cold-dwell fatigue in Ti alloys. Int. J. Plast. 23, 1061–1083. https://doi.org/10.1016/j.ijplas.2006.10.013

Evans, A.G., Rawlings, R.D., 1969. The Thermally Activated Deformation of Crystalline Materials. Phys. Status Solidi 34, 9–31. https://doi.org/10.1002/pssb.19690340102

Feather, W.G., Lim, H., Knezevic, M., 2020. A numerical study into element type and mesh resolution for crystal plasticity finite element modeling of explicit grain structures. Comput. Mech. 23–30. https://doi.org/10.1007/s00466-020-01918-x

Fisher, E.S., Renken, C.J., 1964. Single-crystal elastic moduli and the hcp → bcc transformation in Ti, Zr, and Hf. Phys. Rev. 135. https://doi.org/10.1103/PhysRev.135.A482

Gong, J., Benjamin Britton, T., Cuddihy, M.A., Dunne, F.P.E., Wilkinson, A.J., 2015. (a) Prismatic, (a) basal, and (c+a) slip strengths of commercially pure Zr by micro-cantilever tests. Acta Mater. 96, 249–257. https://doi.org/10.1016/j.actamat.2015.06.020

Gurao, N.P., Akhiani, H., Szpunar, J.A., 2014. Pilgering of Zircaloy-4: Experiments and simulations. J. Nucl. Mater. 453, 158–168. https://doi.org/10.1016/j.jnucmat.2014.06.047

Hayes, T.A., Kassner, M.E., 2017. Creep of zirconium and zirconium alloys. Miner. Met. Mater. Ser. Part F2, 103–114. https://doi.org/10.1007/978-3-319-51097-2_9

Hirth, J.P., Nix, W.D., 1969. An Analysis of the Thermodynamics of Dislocation Glide. Phys. Status Solidi 35, 177–188. https://doi.org/10.1002/pssb.19690350116

Holmes, J.J., 1964. The activation energies for creep of zircaloy-2. J. Nucl. Mater. 13, 137–141. https://doi.org/10.1016/0022-3115(64)90035-2

Jun, T.S., Armstrong, D.E.J., Britton, T.B., 2016a. A nanoindentation investigation of local strain rate sensitivity in dual-phase Ti alloys. J. Alloys Compd. 672, 282–291. https://doi.org/10.1016/j.jallcom.2016.02.146

Jun, T.S., Zhang, Z., Sernicola, G., Dunne, F.P.E.E., Britton, T.B., 2016b. Local strain rate sensitivity of single α phase within a dual-phase Ti alloy. Acta Mater. 107, 298–309. https://doi.org/10.1016/j.actamat.2016.01.057

Klepaczko, J.R., Chiem, C.Y., 1986. On rate sensitivity of f.c.c. metals, instantaneous rate sensitivity and rate sensitivity of strain hardening. J. Mech. Phys. Solids 34, 29–54. https://doi.org/10.1016/0022-5096(86)90004-9

Knezevic, M., Drach, B., Ardeljan, M., Beyerlein, I.J., 2014. Three dimensional predictions of grain scale plasticity and grain boundaries using crystal plasticity finite element models. Comput. Methods Appl. Mech. Eng. 277, 239–259. https://doi.org/10.1016/j.cma.2014.05.003

Knezevic, M., Zecevic, M., Beyerlein, I.J., Bingert, J.F., McCabe, R.J., 2015. Strain rate and temperature effects on the selection of primary and secondary slip and twinning systems in HCP Zr. Acta Mater. 88, 55–73. https://doi.org/10.1016/j.actamat.2015.01.037

Knezevic, M., Zecevic, M., Beyerlein, I.J., Lebensohn, R.A., 2016. A numerical procedure enabling accurate descriptions of strain rate-sensitive flow of polycrystals within crystal visco-plasticity theory. Comput. Methods Appl. Mech. Eng. 308, 468–482. https://doi.org/10.1016/j.cma.2016.05.025

Kombaiah, B., Murty, K.L., 2015. Dislocation cross-slip controlled creep in Zircaloy-4 at high stresses. Mater. Sci. Eng. A 623, 114–123. https://doi.org/10.1016/j.msea.2014.11.040

Kumar, N., Alomari, A., Murty, K.L., 2018. Understanding thermally activated plastic deformation behavior of Zircaloy-4. J. Nucl. Mater. 504, 41–49. https://doi.org/10.1016/j.jnucmat.2018.03.031

Lee, K.W., Kim, S.K., Kim, K.T., Hong, S.I., 2001. Ductility and strain rate sensitivity of Zircaloy-4 nuclear fuel claddings. J. Nucl. Mater. 295, 21–26. https://doi.org/10.1016/S0022-3115(01)00509-8

Link, T.M., Koss, D.A., Motta, A.T., 1998. Failure of Zircaloy cladding under transverse plane-strain deformation. Nucl. Eng. Des. 186, 379–394. https://doi.org/10.1016/S0029-5493(98)00284-2

Matsuo, Y., 1987. Thermal creep of zircaloy-4 cladding under internal pressure. J. Nucl. Sci. Technol. 24, 111–119. https://doi.org/10.1080/18811248.1987.9735783

Mehrotra, B.N., Tangri, K., 1980. High tenperature (600-800 degree C) thermally activated deformation behaviour of alpha -zircaloy-4-oxygen alloys. Acta Metall. 28, 1385–1394. https://doi.org/10.1016/0001-6160(80)90007-3

Northwood, D.O., London, I.M., Bähen, L.E., 1975. Elastic constants of zirconium alloys. J. Nucl. Mater. 55, 299–310. https://doi.org/10.1016/0022-3115(75)90071-9

Nye, J.F., 1953. Some geometrical relations in dislocated crystals. Acta Metall. 1, 153–162. https://doi.org/10.1016/0001-6160(53)90054-6

Quey, R., Dawson, P.R., Barbe, F., 2011. Large-scale 3D random polycrystals for the finite element method: Generation, meshing and remeshing. Comput. Methods Appl. Mech. Eng. 200, 1729–1745. https://doi.org/10.1016/j.cma.2011.01.002

Shi, S.Q., Shek, G.K., Puls, M.P., 1995. Hydrogen concentration limit and critical temperatures for delayed hydride cracking in zirconium alloys. J. Nucl. Mater. 218, 189–201. https://doi.org/10.1016/0022-3115(94)00405-6

Skippon, T., Cochrane, C., Daymond, M.R., 2019. Crystallographic orientation sensitive measurement of strain rate sensitivity in Zircaloy-2 via synchrotron X-ray diffraction. Int. J. Plast. 113, 1–17. https://doi.org/10.1016/j.ijplas.2018.09.003

Sun Ig Hong, Woo Seog Ryu, Chang Saeng Rim, 1984a. Thermally activated deformation of Zircaloy-4. J. Nucl. Mater. 120, 1–5. https://doi.org/10.1016/0022-3115(84)90165-X

Sun Ig Hong, Woo Seog Ryu, Chang Saeng Rim, 1984b. Thermally activated deformation of Zircaloy-4. J. Nucl. Mater. 120, 1–5. https://doi.org/10.1016/0022-3115(84)90165-X

Taylor, G.I., 1934. The mechanism of plastic deformation of crystals. Part I.—Theoretical. Proc. R. Soc. London. Ser. A, Contain. Pap. a Math. Phys. Character 145, 388–404.

https://doi.org/10.1098/rspa.1934.0107

Tong, V.S., Britton, T. Ben, 2017. Formation of very large 'blocky alpha' grains in Zircaloy-4. Acta Mater. 129, 510–520. https://doi.org/10.1016/j.actamat.2017.03.002

Vasilev, E., Zecevic, M., McCabe, R.J., Knezevic, M., 2020. Experimental verification of a crystal plasticity-based simulation framework for predicting microstructure and geometric shape changes: Application to bending and Taylor impact testing of Zr. Int. J. Impact Eng. 144, 103655. https://doi.org/10.1016/j.ijimpeng.2020.103655

Wan, W., 2020. Microstructurally-Sensitive Short Crack Growth in Zircaloy-4. Imperial College London.

Wang, S., Giuliani, F., Britton, T. Ben, 2019. Variable temperature micropillar compression to reveal <a> basal slip properties of Zircaloy-4. Scr. Mater. 162, 451–455. https://doi.org/10.1016/j.scriptamat.2018.12.014

Wiesinger, F.W., Lewis, D.M., Azzarto, F.J., Baldwin, E.E., 1968. Unirradiated, in-pile and post-irradiation low strain rate tensile properties of zircaloy-4 30, 208–218.

Wilson, D., Dunne, F.P.E., 2019. A mechanistic modelling methodology for microstructure-sensitive fatigue crack growth. J. Mech. Phys. Solids 124, 827–848. https://doi.org/10.1016/j.jmps.2018.11.023

Wilson, D., Wan, W., Dunne, F.P.E., 2019. Microstructurally-sensitive fatigue crack growth in HCP, BCC and FCC polycrystals. J. Mech. Phys. Solids 126, 204–225. https://doi.org/10.1016/j.jmps.2019.02.012

Xiong, Y., Karamched, P., Nguyen, C.T., Collins, D.M., Magazzeni, C.M., Tarleton, E., Wilkinson, A.J., 2020. Cold creep of titanium: Analysis of stress relaxation using synchrotron diffraction and crystal plasticity simulations. Acta Mater. 199, 561–577. https://doi.org/10.1016/j.actamat.2020.08.010

Zecevic, M., Beyerlein, I.J., McCabe, R.J., McWilliams, B.A., Knezevic, M., 2016. Transitioning rate sensitivities across multiple length scales: Microstructure-property relationships in the Taylor cylinder impact test on zirconium. Int. J. Plast. 84, 138–159. https://doi.org/10.1016/j.ijplas.2016.05.005

Zhang, Z., Dunne, F.P.E.E., 2017. Microstructural heterogeneity in rate-dependent plasticity of multiphase titanium alloys. J. Mech. Phys. Solids 103, 199–220. https://doi.org/10.1016/j.jmps.2017.03.012

Zhang, Z., Jun, T.S., Britton, T.B., Dunne, F.P.E., 2016a. Intrinsic anisotropy of strain rate sensitivity in single crystal alpha titanium. Acta Mater. 118, 317–330. https://doi.org/10.1016/j.actamat.2016.07.044

Zhang, Z., Jun, T.S., Britton, T.B., Dunne, F.P.E., 2016b. Determination of Ti-6242 α and β slip properties using micro-pillar test and computational crystal plasticity. J. Mech. Phys. Solids

95, 393–410. https://doi.org/10.1016/j.jmps.2016.06.007

Zheng, Z., Balint, D.S., Dunne, F.P.E., 2016. Rate sensitivity in discrete dislocation plasticity in hexagonal close- packed crystals. Acta Mater. 107, 17–26. https://doi.org/10.1016/j.actamat.2016.01.035

**Appendix A. Boundary conditions of truncated model**

The truncated sub-model is 3D and a model depth of 50 μm is chosen to extrude along out-of-plane direction with 5 elements in thickness. 3-5 elements are analysed to obtain the good prediction of stress distribution on the surface of the sample. Shown in Figure A1, a non-uniform stress distribution as a function of the y coordinate is applied at both side surface of the truncated model, $\sigma_x = 204 - 0.146y$. $U_x = 0$ is applied at the middle line of the whole model; $U_y = 0$ is applied at both bottom edges and $U_z = 0$ is applied at both back edge of the model. The non-uniform stress applied at both sides is the key to achieve the stress distribution matching between the sub-model and whole model. A uniform stress distribution would change the curvature of $\sigma_{xx}$ distribution curve due to inaccurate stress calculated from the three-point bending test. For the sub-model truncated from sample under three-point bending, $U_x$ should not be fixed at the bottom edge, leading to unrealistic plastic strain accumulation at the corner of the model, which would lower the stress level along A-A' path. And the top middle line is $U_x = 0$ to avoid the X-displacement instability and to create a near $U_x$ free condition. If Z-axis displacement is fixed at the back surface, it leads to over-constrained condition and the stress level along A-A' path would be over-predicted.

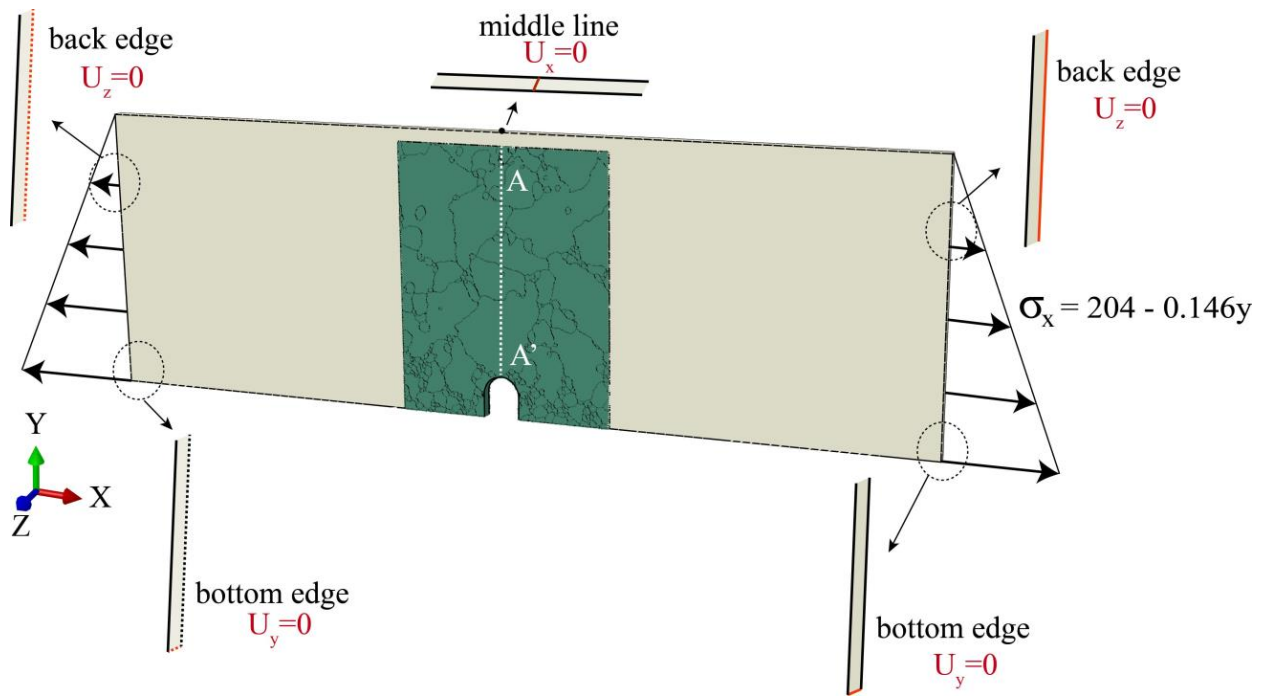

*Figure A1 shows detailed boundary conditions for the truncated sub-model of three-point bending test.*

**Appendix B**. **Mesh sensitivity study for RVE model**

Here, the sensitivities of mesh-type, element type and the mesh resolution on the polycrystalline responses are shown in Figure A2. Three types of meshes are considered including the conformal tetrahedral mesh, the non-conformal hexahedral medium and dense mesh. Linear and quadratic element types are discussed in both tet and hex meshes. Here, the C3D20 medium, C3D20 dense, C3D10 and the C3D8 dense give consistent polycrystalline responses. The nonconformal C3D8 mesh leads to slight deviation because of the increased local plastic deformation due to the staircase effect (Choi et al., 2012; Feather et al., 2020). Whereas, the non-conformal C3D20 (dense or medium) mesh and C3D8 dense mesh lead to similar polycrystalline responses compared with the conformal C3D10 mesh, which demonstrates that the non-conformal mesh could give valid results when suitable element or integration point number are considered in the model, which has been validated in modelling of Nickel-based alloy (Chen, 2018).

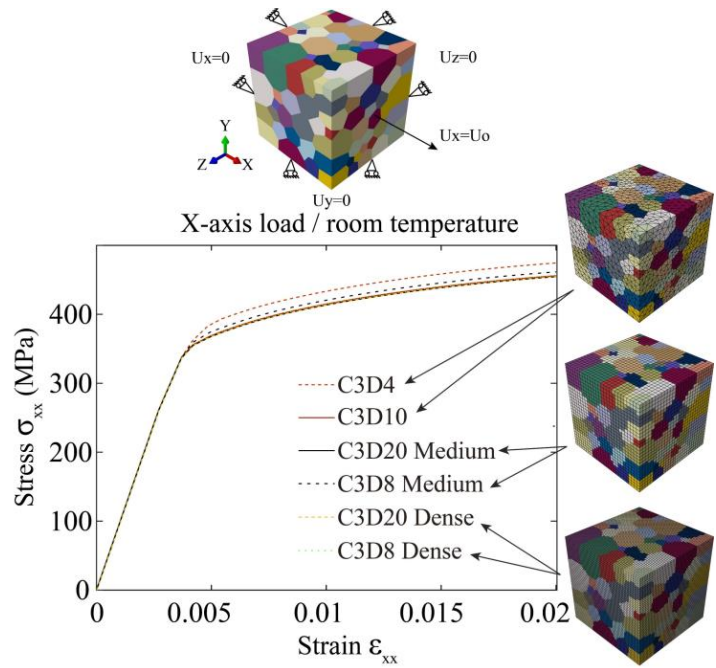

*Figure A2 shows polycrystalline stress-strain responses under different mesh densities and element types.*

It should be noted that different from (Choi et al., 2012; Feather et al., 2020), the length-scale effect has been considered here and the geometrical-necessary dislocation density is incorporated in the CPFEM framework, which would lead to different conclusions to the non-conformal mesh. The non-conformal C3D20 medium mesh used in the manuscript is good enough to capture the polycrystalline stress-strain curve. In addition, the conformal C3D4 mesh leads to much higher stress levels due to the enhanced stiffness of the C3D4, which shows similar results in the mentioned paper (Feather et al., 2020). This would be relaxed by using a stabilised tetrahedral element (Cheng et al., 2016). Here, we use the NEPER software (Quey et al., 2011) to generate the polycrystalline meshes. The three-dimensional modelling of conformal grain structure with hexahedral and tetrahedral meshes are acknowledged (Barrett et al., 2018; Feather et al., 2020; Knezevic et al., 2014), in addition to the automated procedure on the Voxel model from DREAM3D to conformal grain boundary meshes (Barrett et al., 2018).

The local grain response is now discussed for the mentioned mesh type and element type in Figure A3. The specific grain selected has dominant active slip systems of basal and pyramidal <a>. It is demonstrated that the non-conformal mesh of C3D20 (dense or medium) and C3D8 dense could reach similar local grain responses as the conformal mesh of C3D10. The slight deviation of stress-strain curve is observed for C3D8 medium, and a much higher stress elevation is obtained for C3D4 mesh. This indicates that the mesh type has a higher influence over the local grain responses while using the medium mesh density.

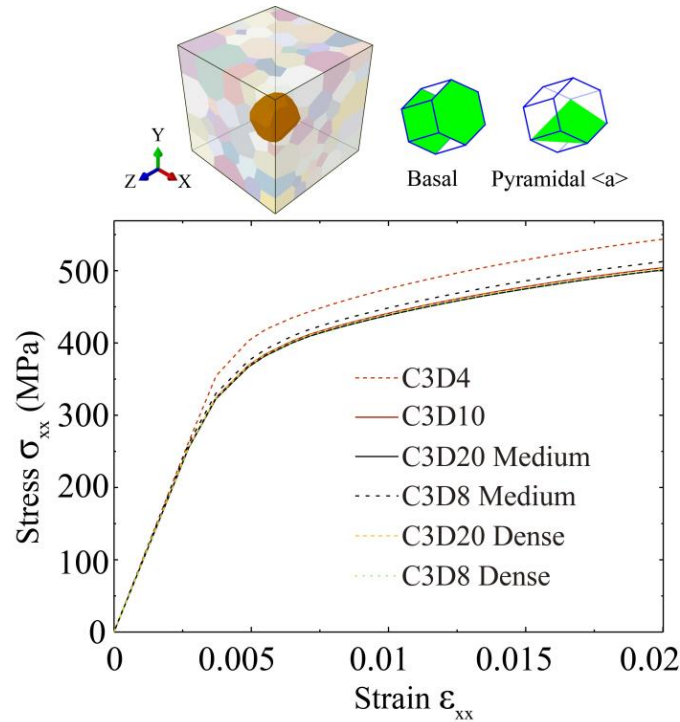

*Figure A3 shows the inner grain stress-strain responses under different mesh densities and element types with predominant activated slip system of basal and pyramidal <a>.*

In conclusion, C3D20 element type and the medium mesh density, used in the RVE modelling of as-received Zircaloy-4, gives reasonable local-grain and polycrystalline responses, compared with that of the conformal element type and denser mesh.